\documentclass[aps,prb,nobibnotes,reprint,twocolumn,superscriptaddress]{revtex4-1}

\usepackage{amsmath, amssymb, amsfonts}
\usepackage{array, booktabs, tabularx}
\usepackage{xcolor, graphicx, tikz}
\usepackage[caption=false]{subfig}
\usepackage{hyperref}  
\usepackage{braket}
\usepackage{bm}
\usepackage{cancel}
\usetikzlibrary{matrix,decorations.pathreplacing,arrows.meta,positioning,calc}
\usetikzlibrary{backgrounds}

\hypersetup{colorlinks = True, allcolors = {blue}}

\newcommand{\BC}[1]{\hat{a}_{#1}^\dagger}
\newcommand{\BA}[1]{\hat{a}_{#1}}
\newcommand{\BN}[1]{\hat{n}_{#1}}

\newcommand{\KetBra}[2]{\ket{#1} \bra{#2}}
\newcommand{\Proj}[2]{\mathbb{P}_{#2}^{(#1)}}
\def\UniER{ U_{ER} }

\def\EYE{ \mathbb{I} }

\newcommand{\ComCom}[1]{\mathcal{O}(#1)}

\newcommand{\Eq}[1]{Eq.~({#1})}
\newcommand{\Fig}[1]{Figure~{#1}}

\newcommand{\Sec}[1]{Section~{#1}}

\newcommand{\Reference}[1]{Ref.~{#1}}

\begin{document}


\title{Solving Constrained Optimization Problems Using \\ Hybrid Qubit-Qumode Quantum Devices}

\author{Rishab Dutta}
\affiliation{Department of Chemistry, Yale University, New Haven, CT, USA 06520}

\author{Brandon Allen}
\thanks{These authors contributed equally to this work.}
\affiliation{Department of Chemistry, Yale University, New Haven, CT, USA 06520}

\author{Chuzhi Xu}
\thanks{These authors contributed equally to this work.}
\affiliation{Department of Chemistry, Yale University, New Haven, CT, USA 06520}

\author{Nam P. Vu}
\affiliation{Department of Chemistry, Yale University, New Haven, CT, USA 06520}
\affiliation{Department of Electrical Engineering and Computer Science, Massachusetts Institute of Technology, Cambridge, MA, USA 02139}
\affiliation{Research Laboratory of Electronics, Massachusetts Institute of Technology, Cambridge, MA, USA 02139}

\author{Kun Liu}
\affiliation{Department of Computer Science, Yale University, New Haven, CT, USA 06520}

\author{Fei Miao}
\affiliation{School of Computing, University of Connecticut, Storrs, CT, USA 06269}

\author{Bing Wang}
\affiliation{School of Computing, University of Connecticut, Storrs, CT, USA 06269}

\author{Amit Surana}
\affiliation{RTX Technology Research Center, East Hartford, CT, USA 06118}

\author{Chen Wang}
\affiliation{Department of Physics, University of Massachusetts-Amherst, Amherst, MA, USA 01003}

\author{Yongshan Ding}
\affiliation{Department of Computer Science, Yale University, New Haven, CT, USA 06520}
\affiliation{Department of Applied Physics, Yale University, New Haven, CT, USA 06520}
\affiliation{Yale Quantum Institute, Yale University, New Haven, CT, USA 06511}

\author{Victor S. Batista}
\affiliation{Department of Chemistry, Yale University, New Haven, CT, USA 06520}
\affiliation{Yale Quantum Institute, Yale University, New Haven, CT, USA 06511}
\email{victor.batista@yale.edu}


\begin{abstract}

Variational Quantum Algorithms (VQAs) provide a promising framework for tackling complex optimization problems on near-term quantum hardware. 
Here, we demonstrate that hybrid qubit--qumode quantum devices offer an efficient route to solving Quadratic Unconstrained Binary Optimization (QUBO) problems using the Echoed Conditional Displacement Variational Quantum Eigensolver (ECD-VQE). 
Leveraging circuit quantum electrodynamics (cQED) architectures, we encode QUBO instances across multiple qumodes weakly coupled to a single qubit and extract binary solutions directly from photon-number measurements. 
We apply ECD-VQE to the Binary Knapsack Problem and show that it outperforms the Quantum Approximate Optimization Algorithm (QAOA) implemented on conventional qubit circuits, achieving higher-quality solutions with dramatically fewer resources. We also demonstrate that ECD-VQE can be extended to chemically motivated tasks such as active-space selection for multireference electronic structure methods. 
These results highlight the utility of hybrid qubit-qumode platforms for a broad class of NP-hard and chemistry-related optimization problems, and demonstrate that variational ECD ansatz can realize expressive state preparation with significantly shallower circuits than qubit-only architectures, positioning qubit-qumode gates as compelling candidates for constrained optimization in early fault-tolerant quantum computing.

\end{abstract}


\maketitle


\section{Introduction} 
\label{sec: intro}

Variational Quantum Algorithms (VQAs) represent a powerful class of hybrid quantum--classical methods designed to solve optimization problems on near-term and early fault-tolerant quantum devices. \cite{Mcclean2016theory,Cerezo2021variational,Tilly2022variational,Callison2022hybrid, delgado2025quantum}
VQAs address such problems by encoding them into a Hamiltonian and iteratively minimizing its expectation value through classical optimization over a parametrized quantum ansatz. 
Their broad applicability has been demonstrated across domains including molecular property prediction, \cite{Yawata2022qubo,Ajagekar2023molecular}
RNA folding, \cite{Zaborniak2022qubo}
protein--ligand docking, \cite{Yanagisawa2024qubo}
machine learning, \cite{Date2021qubo}
structural design, \cite{Matsumori2022application}
vehicle routing, \cite{Feld2019hybrid}
and capital budgeting. \cite{Laughhunn1970quadratic}
These problems naturally admit formulations as quadratic unconstrained binary optimization (QUBO) models \cite{Glover2018tutorial,Glover2022applications,Lucas2014ising,Mucke2019learning}, which can be solved using VQAs. 
Indeed, many chemical problems can be formulated as a QUBO problem, and as discussed later in this paper, one example is finding the optimal orbitals for electronic structure calculations.\cite{xia2017electronic}

\begin{figure*}[t!]
\centering
\includegraphics[width=0.9\textwidth]{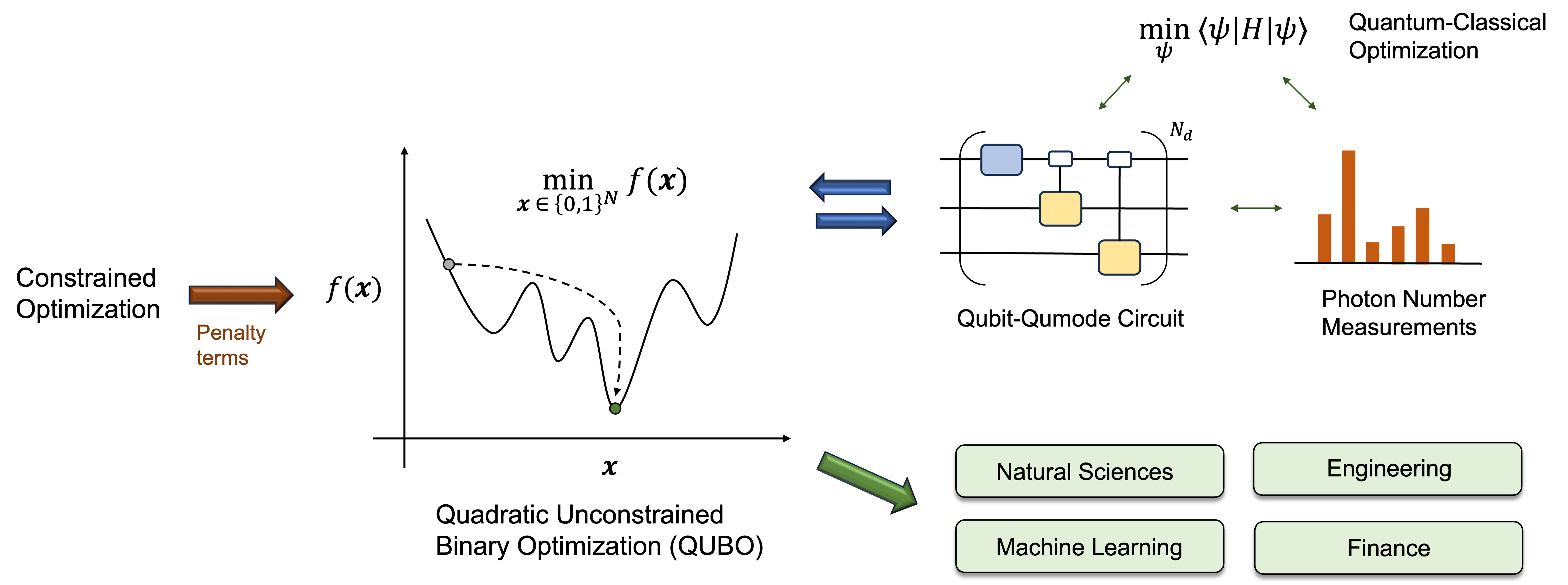}
\caption{
A constrained optimization problem is reformulated as a QUBO instance and solved using a qubit--qumode device. 
The problem is encoded into a qubit--qumode Hamiltonian whose ground state corresponds to the optimal solution. 
This ground state is approximated via variational optimization of the circuit parameters, and the solution is read out through photon-number measurements on the qumodes.}
\label{fig: overview}
\end{figure*}

Two well-known VQAs, the Variational Quantum Eigensolver (VQE), \cite{Peruzzo2014,Cerezo2021variational,Tilly2022variational,grimsley2019adaptive, michel2023blueprint, sherbert2025parametrization}
and the Quantum Approximate Optimization Algorithm (QAOA), \cite{Farhi2014quantum,Blekos2024review}
illustrate the diversity of variational approaches. 
VQE, originally developed for quantum chemistry, often employs either general-purpose ansatze or chemically motivated constructions such as unitary coupled cluster (UCC). \cite{Taube2006new,Anand2022quantum}
In contrast, QAOA constructs its ansatz directly from the problem Hamiltonian and a noncommuting mixing operator, making it particularly suitable for constrained combinatorial optimization. 
Despite their success, both methods can experience trainability challenges including barren plateaus and complex variational landscapes. \cite{Willsch2020benchmarking,Rajakumar2024trainability,Singhal2024performance,muller2025limitations}
These limitations motivate the exploration of improved ansatz architectures, alternative variational principles such as quantum imaginary-time evolution, \cite{Mcardle2019,Kyaw2023boosting} 
and new quantum platforms along with their native gates. \cite{Dutta2024perspective,Dutta2024EST}

In this paper, we focus on quantum harmonic oscillators, also known as qumodes, as the building block of quantum computing resources. \cite{Weedbrook2012} 
Qumodes offer, at least in principle, infinite-dimensional Hilbert spaces, which have the potential to go beyond conventional qubits in expressivity and resourcefulness. 
This flexibility has inspired optimization algorithms \cite{Verdon2019quantum,Enomoto2023continuous,Khosravi2023mixed,Chandarana2024photonic}
tailored to photonic platforms such as integrated photonics and Gaussian boson sampling. \cite{OBrien2009photonic,Wang2020integrated,Arrazola2021quantum}
However, photonic architectures face challenges in implementing strong non-Gaussian operations, whereas circuit quantum electrodynamics (cQED) enables efficient non-Gaussian gates through strong qubit-qumode coupling. \cite{Blais2021,Copetudo2024}
This makes cQED-based hybrid architectures a promising path toward early fault-tolerant quantum computing, with recent demonstrations of quantum error correction beyond break-even for both logical qubits \cite{Sivak2023real} and qudits. \cite{brock2025quantum}

Hybrid cQED systems, built from superconducting cavities coupled to transmon qubits, \cite{Blais2021,Copetudo2024}
enable high-fidelity universal gate sets \cite{Krastanov2015,Eickbusch2022,Diringer2024cnot,You2024Crosstalk,Zhang2024energy}
and exhibit intrinsic noise resilience. \cite{Sivak2023real,brock2025quantum}
Importantly, they support expressive ansatze composed of native operations such as echoed conditional displacement (ECD) gates with qubit rotations, \cite{Eickbusch2022,You2024Crosstalk,Zhang2024energy}
SNAP gates with Gaussian operations, \cite{Krastanov2015,liu2026hybrid}
and conditional-not displacement gates. \cite{Diringer2024cnot}
These hybrid operations allow efficient exploration of large Hilbert spaces, \cite{Zhang2024energy} 
often with dramatically fewer resources than qubit-only circuits, \cite{Wang2020vibronic,liu2026hybrid}
making them a compelling architecture for near-term quantum applications.

In this work, we show how to leverage hybrid qubit-qumode circuits to solve various Quadratic Unconstrained Binary Optimization (QUBO) problems as outlined in \Fig{\ref{fig: overview}}, which fall under the NP-hard class relevant across logistics, scheduling, finance, materials discovery, drug development, and
machine learning. \cite{Glover2018tutorial,Glover2022applications,Lucas2014ising,Mucke2019learning}
Importantly, we demonstrate how to solve a specific quantum chemistry problem related to molecular electronic structure by translating it into a QUBO problem and then solving it using our approach. 
We introduce QUBO formulations with both soft and hard constraints using ECD-VQE, a variational quantum algorithm built from echoed conditional displacement gates and qubit rotations. \cite{Quintero2021characterizing,Bontekoe2023translating}
ECD-VQE encodes QUBO Hamiltonians across multiple qumodes coupled to a single qubit, yielding compact and highly expressive representations of the optimization landscape. 
The circuit parameters are optimized classically, and binary decision variables are extracted from photon-number measurements on the qumodes.

We benchmark ECD-VQE on constrained optimization problems and demonstrate that it outperforms the Quantum Approximate Optimization Algorithm (QAOA) on the Binary Knapsack Problem (BKP), \cite{Martello1990knapsack}
achieving higher solution quality and more favorable variational landscapes. 
For the BKP, ECD-VQE consistently identifies the optimal solution with $100\%$ probability in around 80 iterations, while QAOA fails to exceed a $12\%$ success probability even with the same number of parameters, more iterations, and 50 parameter initializations. 
Notably, ECD-VQE achieves this performance using only 10 native ECD gates with two qumodes and one qubit; an equivalent qubit-only implementation would require approximately 930 CNOT gates on a seven-qubit circuit (Sec.~\ref{sec: application}), underscoring the resource efficiency of the hybrid approach.

The remainder of the paper is organized as follows. 
In \Sec{\ref{sec: methods}}, we describe the mapping of constrained optimization problems onto hybrid qubit--qumode Hamiltonians and detail the ECD-VQE circuit architecture. 
Sections \ref{sec: application} and \ref{sec: chem_qubo} apply the method to benchmark optimization problems and to chemically motivated tasks such as active-space selection. 
Finally, \Sec{\ref{sec: final}} summarizes our findings and outlines directions for future work.
Additionally, we evaluate performance under simulated noise in Appendix \ref{app: noise}.


\section{Methods} \label{sec: methods}

In this section, we review how to represent a constrained optimization problem in terms of a qubit Hamiltonian before discussing how to implement it on a qubit-qumode device.
Then, we introduce a variational approach to find the optimal solution as the ground state of the qubit Hamiltonian. 

\subsection{Qubit Hamiltonian} \label{sec: bkp_qubo_qubit}

Let us review how to transform a constrained optimization problem into a QUBO form for the binary knapsack problem (BKP), a fundamental integer programming problem in combinatorial optimization and operations research. \cite{Martello1990knapsack}
It can be defined as 
\begin{subequations} \label{eq: binary_knapsack}
\begin{align}
\max_{\mathbf{x}} \:
V 
&= \sum_{j = 0}^{N_0 - 1} \: v_j \: x_j, \quad 
x_j \in \{ 0, 1 \},
\\
\text{subject to} \quad 
W_0 (\mathbf{x}) 
&= \sum_{j = 0}^{N_0 - 1} \: w_j \: x_j \leq W,
\end{align}
\end{subequations}
where $W$ is the total weight capacity of a knapsack, $N_0$ is the number of items available, $\{ v_j \}$ are the item values, and $\{ w_j \}$ are the item weights. 

The BKP optimization problemd defined in \Eq{\ref{eq: binary_knapsack}} involves maximizing $V$ with hard constraints requiring total weight to be smaller than $W$.  
It is also called the 0-1 knapsack problem, and is known to belong to the NP-hard computational complexity class. \cite{Martello1997upper,Martello1999dynamic,Pisinger2005hard}
Many real-world optimization problems can be represented as a BKP, \cite{Salkin1975knapsack} 
including optimization tasks for molecular drug discovery. \cite{Yevseyeva2019application}

The constrained optimization of \Eq{\ref{eq: binary_knapsack}} can be transformed to an unconstrained problem by introducing auxiliary binary variables $\{ y_j \}$ to have a QUBO representation: \cite{Quintero2021characterizing,Bontekoe2023translating}
\begin{equation} \label{eq: binary_knapsack_qubo}
\min_{\mathbf{x}, \mathbf{y}} E 
= - V (\mathbf{x})
+ \lambda \: \Big[ 
W - W_0 (\mathbf{x}) 
- \sum_{j = 0}^{N_1 - 1} \: 2^j \: y_j \Big]^2,
\end{equation}
where $\lambda$ is the quadratic penalty weight and 
the sum of $ N_1 = \lceil \log_2 (W + 1) \rceil $
terms. 
We can now write the cost function $E (\mathbf{x})$ as a function of $N_0 + N_1$ binary variables $\{ x_j \}$, including the auxiliary variables:
\begin{align} \label{eq: binary_knapsack_unconstrained}
\min_{\mathbf{x}} E 
&= - \sum_{j = 0}^{N_0 - 1} \: v_j \: x_j 
+ \lambda \: \Big[ W 
- \sum_{j = 0}^{N_0 - 1} \: w_j \: x_j \nonumber
\\
&- \sum_{j = N_0}^{N_0 + N_1 - 1} 2^{j - N_0} \: x_j \Big]^2.
\end{align}
This cost function can now be mapped to a qubit Hamiltonian $H_Q$ of 
$N = N_0 + N_1$ qubits by substituting each binary variable, as follows:
$x_j \mapsto \frac{1}{2} \: ( \EYE_j - Z_j ) $, where $\EYE$ and $Z$ are the identity and Pauli-Z operators, while the subscript index represents the qubit site. 
The mapping from binary variable to qubit operator can be easily justified by noting that the eigenstates of the $Z$ operator are the qubit basis states $\ket{0}$ and $\ket{1}$ with eigenvalues $+1$ and $-1$, respectively. 
The multi-qubit Hamiltonian can then be written, as follows: 
\begin{align} \label{eq: bkp_qubit_ham}
H_Q
&= - \sum_{j = 0}^{N_0 - 1} \: \frac{v_j}{2} \: ( \EYE_j - Z_j )
+ \lambda \: \Big[ W 
- \sum_{j = 0}^{N_0 - 1} \: \frac{w_j}{2} \: ( \EYE_j - Z_j ) \nonumber
\\
&- \sum_{j = N_0}^{N_0 + N_1 - 1} 2^{j - N_0 - 1} \: ( \EYE_j - Z_j ) \Big]^2.
\end{align}
The binary string $\mathbf{x}^*$ representing the optimal solution of \Eq{\ref{eq: binary_knapsack_unconstrained}} is now encoded into a tensor product of $N$ computational basis states 
\begin{equation}
\ket{\psi}
= \ket{x_0^*} \otimes \cdots \otimes \ket{x_{N - 1}^*},
\end{equation}
and is the ground state of $H_Q$.
Similar to the BKP problem discussed above, any constrained optimization problem can be represented by a Hamiltonian $H_Q$ of the form defined in \Eq{\ref{eq: bkp_qubit_ham}} by representing the constraints by auxiliary variables and designing a quadratic penalty function. 

\subsection{Hilbert space mapping} \label{sec: mapping}

Let us define a $N$-qubit Hamiltonian $H_D$ below that can be written as a linear combination of terms, each consisting of only identity and Pauli-Z operators  
\begin{equation} \label{eq: diag_ham_qubit}
H_D 
= \sum_{\mu = 1}^{N_H} \: g_\mu \: 
\sigma_1^{(\mu)} \otimes \cdots \otimes 
\sigma_{N}^{(\mu)} 
= \sum_{\mu = 1}^{N_H} \: g_\mu \: 
\mathcal{D}_{N}^{(\mu)},
\end{equation}
where $\sigma_j = \EYE, Z$, the Hamiltonian coefficients $\{ g_\mu \}$ are known, and 
the number of terms $N_H$ is assumed to be a computationally manageable finite number. 
For example, $H_Q$ defined in \Eq{\ref{eq: bkp_qubit_ham}} consists of 
$\ComCom{ N_0^2 + N_1^2 + N_0 N_1 }$ terms.  
Each of the 
$ \mathcal{D}_{N}^{(\mu)} $ terms is a diagonal operator since 
\begin{subequations} \label{eq: diag_ops_qubit}
\begin{align}
\EYE 
&= \KetBra{0}{0} + \KetBra{1}{1},
\\
Z 
&= \KetBra{0}{0} - \KetBra{1}{1}. 
\end{align}    
\end{subequations} 
Our goal is to compute 
$ \braket{\psi| \: \mathcal{D}_{N}^{(\mu)} \: | \psi} $ 
for a given state using a combination of Pauli-Z measurements on a qubit and photon number measurements on qumodes. 
The photon number measurements compute the probabilities of finding the discrete Fock states $\{ \ket{n} \}_{n\in\mathbb{N}}$ of a quantum harmonic oscillator or the number of photons in an optical mode, and are the eigenstates of the bosonic number operator, 
$ \BN{} \ket{n} 
= \BC{} \BA{} \ket{n} = n \ket{n} $, 
where $ \hat{a}^\dagger, \hat{a} $ are bosonic creation and annihilation operators, respectively.
For a realistic setup, the maximum number of photons can be set to a finite integer $L - 1$, where $L$ is called the Fock cutoff. 
From the Fock basis perspective, a qumode is thus equivalent to a multilevel generalization of a qubit in $L$ dimensions, also known as a qudit \cite{Wang2020Qudits}.

Let us first discuss the computation of the expectation values in the context of photon number measurements of one qumode for the sake of simplicity. 
The observable $ \braket{\psi| \: \mathcal{D}_{N}^{(\mu)} \: | \psi} $ can be computed from the histogram of all possible binary strings $\ket{q_1, \cdots, q_N}_Q$ from Pauli-Z measurements since
\begin{equation} \label{eq: exp_val_qubit}
\braket{\psi| \: 
Z_{p_1} \cdots Z_{p_N} \: | \psi}
= \sum_{\mathbf{b}} (-1)^{\sum_{i=1}^N b_{p_i}} \: \mathbb{P} (\mathbf{b}),
\end{equation}
where $\{ \mathbf{b} \}$ represent all qubit basis states as bitstrings, 
$\mathbb{P} (\mathbf{b})$ is the probability of measuring the state 
$\ket{\mathbf{b}}$ and 
$ \sum_{i=1}^N b_{p_i} $ is the sum of the bit values at positions 
$ \{ p_1, \cdots, p_N \} $, which determines the sign for each qubit in the Pauli word.
Each of the binary bitstrings of $N$ qubits can be in principle mapped to the Fock space of a single qumode with $L = 2^N$ using the binary mapping 
\begin{equation} \label{eq: binary_map}
\ket{q_1, \cdots, q_N}_Q
\leftrightarrow \ket{n}_B, 
\end{equation}
where $ n = 2^0 \: q_1 + \cdots + 2^{N - 1} \: q_N $.
Thus, the histogram of binary strings can be generated by photon number measurements on a single qumode instead of Pauli-Z measurements on multiple qubits. 
A single qumode Hilbert space with a realistic cutoff $L$ can only handle mapping a few qubits realistically. 
However, we can expand the Hilbert space significantly by working with multiple qumodes, thus allowing the mapping of a large number of qubits with fewer qumodes by a constant factor $\log_2(L)$. 
In this work, we will focus on a hardware setup with one qubit and two qumodes, which can be readily generalized to multiple qumodes. 
Thus, we will explore partitioning the $N$-qubit Hilbert space such that it matches with the combined Hilbert space of one qubit and two qumodes, i.e., $ 2^N = 2 \times L_1 \times L_2 $, where $L_1$ and $L_2$ are the Fock cutoffs for the first and second qumodes, respectively.
Thus, the $N$-qubit state $\ket{q_1, \cdots, q_N}_Q$ Hilbert space can now be partitioned into three pieces and mapped to 
\begin{align} \label{eq: binary_map_qcc}
&\ket{q_1}_Q \otimes \ket{q_2, \cdots, q_{N - j}}_Q \otimes
\ket{q_{N - j + 1}, \cdots, q_N}_Q \nonumber
\\
&\leftrightarrow \ket{q_1}_Q \otimes \ket{n}_B \otimes \ket{m}_B,
\end{align}
where $1 \leq j \leq N - 2$. 
For example, one possible partition for the Hamiltonian $H_Q$ defined in \Eq{\ref{eq: bkp_qubit_ham}} can be that the $N_0$ primary variables are represented by the qubit and the first qumode, whereas the second qumode represents the $N_1$ auxiliary variables.
Let us discuss with a simple example where a one-qubit two-qumode quantum state $\ket{\psi}$ is prepared, followed by Pauli-Z measurement on the qubit and photon number measurements on the two qumodes.
Let us also assume $\ket{\psi}$ is originally representing a five-qubit Hamiltonian, which has been partitioned such that the first qubit remains the same, where the rest of the four qubits are grouped into two equal parts and each mapped to one qumode. 
The measured bitstrings of the first subsystem remain the same, i.e., $\ket{0}$ and $\ket{1}$.
For each of the second and third subsystems, the possible bitstrings can be mapped as 
\begin{subequations}
\begin{align}
\ket{0} \otimes \ket{0}_Q 
&\mapsto \ket{0}_B, \quad 
\ket{0} \otimes \ket{ 1}_Q 
\mapsto \ket{1}_B,  
\\
\ket{1} \otimes \ket{ 0}_Q 
&\mapsto \ket{2}_B, \quad 
\ket{1} \otimes \ket{ 1}_Q 
\mapsto \ket{3}_B.  
\end{align} 
\end{subequations}
For example, the one-qubit two-qumode basis state 
$ \ket{ 1 }_Q \otimes \ket{3}_B \otimes \ket{2}_B $ is same as the five-qubit basis state 
$\ket{1, 1, 1, 1, 0}_Q$.


\begin{figure}[t!]

\includegraphics[width=0.9\columnwidth]{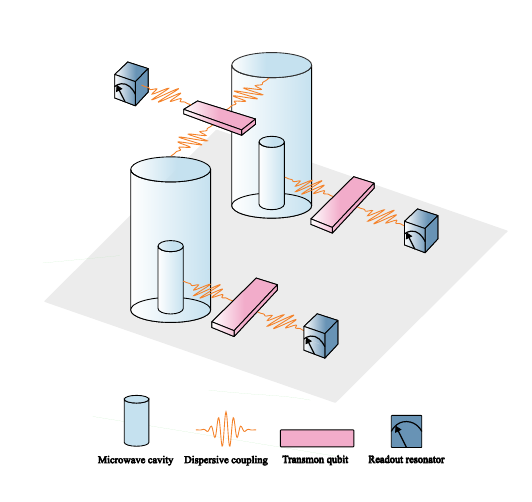}

\caption{
    Schematic to implement quantum nondemolition (QND) approach for photon number measurements for two qumodes with computational measurements for the coupled qubit. 
    Two microwave cavities are dispersively coupled to a coupler transmon qubit with a readout resonator for computational basis measurement of the coupler qubit. 
    Each of the cavities are also coupled to an ancillary transmon qubit with a readout resonator which allows for photon number detection followed by the approach discussed in \Reference{\citenum{Wang2020vibronic}}.
}
\label{fig: photon_num_meas}
\end{figure}


A novel characteristic of our approach is the use of photon number measurements to output positive integers that can be easily converted to binary strings for evaluating the expectation value of the qubit-based Hamiltonian $H_Q$ as defined in \Eq{\ref{eq: diag_ham_qubit}} or $H_D$ of \Eq{\ref{eq: bkp_qubit_ham}}.
Photon number measurements can be implemented using the quantum nondemolition (QND) method, as described in \Reference{\citenum{Wang2020vibronic}}.
The QND measurement relies on the dispersive coupling between the cavity mode and its ancillary transmon qubit. 
In the dispersive regime, the transition frequency of the transmon qubit shifts depending on the photon number in the cavity. 
This shift enables the transmon to encode information about the cavity's photon state. 
A sequence of numerically optimized control pulses is applied to the transmon to extract this information. 
These pulses selectively drive the transmon between its quantum states based on the binary representation of the photon number. 
The transmon state carrying photon number information is probed via a dispersive readout. 
This QND scheme achieves high resolution and fidelity, resolving photon numbers up to 15 in single-shot experiments. \cite{Wang2020vibronic} 
We refer the reader to \Fig{\ref{fig: photon_num_meas}} for a schematic of the hardware setup where photon number measurements on two qumodes is combined with computational basis measurements on a qubit. 

While computing the expectation values of the Hamiltonian in binary basis is a generally effective strategy, we may also rewrite the auxiliary binary variables $\{ y_j \}$ in \Eq{\ref{eq: binary_knapsack_qubo}} in the basis of the photon number cutoff $L$ of the qubits.  As an example, we can use one integer variable $b$,
\begin{equation} 
\min_{\mathbf{x}, \mathbf{y}} E 
= - V (\mathbf{x})
+ \lambda \: \Big[ 
W - W_0 (\mathbf{x}) 
- b \Big]^2,
\end{equation}
such that it has values $b = 0, \cdots, 2^{N_1} - 1$. 
By representing the integer variable to the bosonic number operator $b \mapsto \BN{}$, we can now map the qubit Hamiltonian $H_Q$ defined in \Eq{\ref{eq: bkp_qubit_ham}} to a qubit-qumode Hamiltonian of the form 
\begin{align}
H_Q \mapsto H_{QB}
&= - \sum_{j = 0}^{N_0 - 1} \: \frac{v_j}{2} \: ( \EYE_j - Z_j ) \nonumber
\\
&+ \lambda \: \Big[ W 
- \sum_{j = 0}^{N_0 - 1} \: \frac{w_j}{2} \: ( \EYE_j - Z_j ) 
- \BN{} \Big]^2,
\end{align}
where the number operator $\BN{}$ is assumed to have the Fock cutoff $L = 2^{N_1} - 1$ and the corresponding qumode represents all of the auxiliary variables. 
This ensures that the photon number measurements corresponding to the Hilbert space of the auxiliary variables can be used directly without the integer-to-binary mapping, and to fewer Hamiltonian terms compared to the qubit-only Hamiltonian defined in \Eq{\ref{eq: bkp_qubit_ham}}. 

Finally, it is also possible to represent and evaluate the Hamiltonian directly in the basis of the experimental qubit-qumode device in the projection operator form 
\begin{align} \label{eq: diag_ham_qumode}
H_Q \mapsto H_{QB} \nonumber
&= \sum_{i \in \{Q, B1, B2\} } 
\sum_{n=0}^{L_i} C_n^i \: \Proj{i}{n} \nonumber 
\\
&+ \sum_{ i, j \in \{ Q, B1, B2 \} }^{i\neq j} 
\sum_{n = 0}^{L_i} 
\sum_{m=0}^{L_j} 
C_{n,m}^{i,j} \: \Proj{i}{n} \otimes \Proj{j}{m} 
\end{align}
where $\Proj{i}{n} \equiv \KetBra{n}{n}$ is defined in the Hilbert space of the qubit mode ($i=Q$) of dimension 2 or the bosonic modes ($i=B_1$ or $B_2$) of dimension $L_1$ or $L_2$.
$\{ C_n^i \}$ and $\{ C_{n,m}^{i,j} \}$  can be deduced by rewriting the single- or two-qubit Pauli Z operators in the qubit Hamiltonian \Eq{\ref{eq: bkp_qubit_ham}} following the qubit-to-qumode mapping.  For example, given the mapping Eq. (11), we have ${Z_2 Z_3}=\ket{0}\bra{0}-\ket{1}\bra{1}-\ket{2}\bra{2}+\ket{3}\bra{3}$. 
We note that for a large problem with many qubits and qumodes, the number of terms in the Hamiltonian will only grow quadratically with the number of modes.

\subsection{Variational quantum eigensolver} \label{sec: vqe}

We take a variational quantum eigensolver (VQE) approach to find the approximate ground state of the diagonal Hamiltonian $H_D$ defined in \Eq{\ref{eq: diag_ham_qubit}} by optimizing the following cost function
\begin{equation} \label{eq: vqe}
\min_{\psi} E
= \braket{\psi | \: H_D \: |\psi}
= \sum_{\mu = 1}^{N_H} \: g_\mu 
\braket{\psi | \: \mathcal{D}_{N}^{(\mu)} \: |\psi},
\end{equation}
where $\mathcal{D}_{N}^{(\mu)}$ can be computed using a qubit-qumode device following the steps below. 
\begin{itemize}
    
\item Prepare a normalized trial one-qubit two-qumode state $\ket{\psi}$ using a parameterized quantum circuit. 

\item Generate the histogram of 
$ \ket{q}_Q \otimes \ket{n}_B \otimes \ket{m}_B $ using Pauli-Z measurements on the qubit and photon number measurements on the two qumodes. \cite{Wang2020vibronic} 

\item Compute the expectation value by combining \Eq{\ref{eq: exp_val_qubit}} and \Eq{\ref{eq: binary_map_qcc}}.

\end{itemize} 
The trial state $\ket{\psi}$ can be generated by a parameterized one-qubit two-qumode circuit acting on the vacuum state 
\begin{equation} \label{eq: ansatz}
\ket{\psi (\mathbf{v})} 
= U ( \mathbf{v} ) \: \Big( 
\ket{0}_Q \otimes \ket{0}_C \otimes \ket{0}_C \Big),
\end{equation}
where the vector \textbf{v} represents all the circuit parameters. 
The parameters can be updated by optimizing \Eq{\ref{eq: vqe}} on a classical computer.  The histograms on the qubit-qumode device compute the following probabilities
\begin{equation} \label{eq: probs}
S_{q, n, m}
= | \braket{q, n, m | \psi} |^2, 
\end{equation}
where 
$ q = \{ 0, 1 \} $, 
$ n = \{ 0, 1, \cdots, L_1 - 1 \} $, and 
$ m = \{ 0, 1, \cdots, L_2 - 1 \} $ are the possible basis states.
After repeating the measurement experiments for a finite number of times, a distribution for $\{ S_{q, n, m} \}$ can be generated for computing the expectation values.

The overlaps $\{ S_{q, n, m} \}$ also measure how close the approximate ground state $\ket{\psi}$ --- a state in a superposition of multiple solution strings --- is to the optimal solution string. 
Thus, for the QUBO problems, we only care about the resolution of the distribution generated by the measurements $\{ S_{q, n, m} \}$ instead of how close the trial energy $E$ is to the true ground state energy of the Hamiltonian $H_D$. 
The quantum superposition of $\ket{\psi}$ also highlights the potential advantage of quantum optimization algorithms. 
The parametrized one-qubit two-qumode trial state can be generically represented as   
\begin{equation}
\ket{\psi (\mathbf{v})}
= \sum_{q \in \{ 0, 1 \}} 
\sum_{n = 0}^{L_1 - 1} \sum_{m = 0}^{L_2 - 1}
\lambda_{q, n, m} (\mathbf{v}) \ket{q, n, m},
\end{equation}
and the optimization steps depend on the gradient for the cost function defined in \Eq{\ref{eq: vqe}}  
\begin{equation}
\frac{ \partial E }{ \partial \mathbf{v} }
= \braket{\frac{ \partial \psi }{ \partial \mathbf{v} } | \: H_D \: |\psi} 
+ \braket{ \psi | \: H_D \: | \frac{ \partial \psi }{ \partial \mathbf{v} } }, 
\end{equation}
which in turn updates $\ket{\psi (\mathbf{v})}$ for the next iteration based on \Eq{\ref{eq: ansatz}}.  
This means that at each optimization step, all the combinatorial number of basis state coefficients affect the variational parameters, which in turn update all the basis state coefficients at the same time in the next iteration. 
This can be hard to mimic using a classical distribution that only samples from a subspace of the full Hilbert space. 
\cite{Metropolis1953equation,Hastings1970monte,Mohseni2022ising}
Nevertheless, the true potential of quantum superposition is achieved when the parameterized circuit for $\ket{\psi (\mathbf{v})}$ is sufficiently expressive, which we discuss below in the context of qubit-controlled bosonic qumode gates. 


\begin{figure}[t!]

\includegraphics[width=0.9\columnwidth]{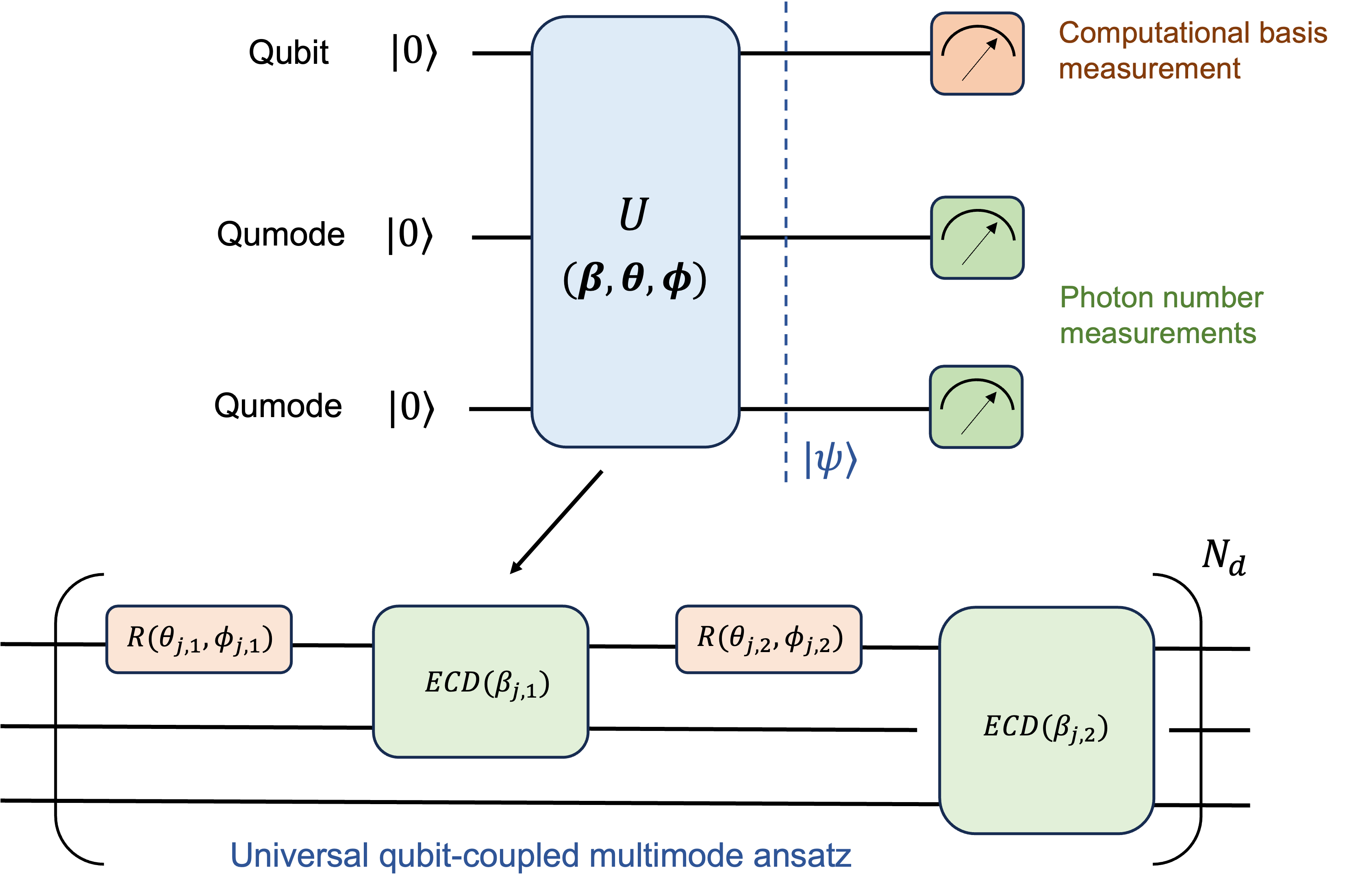}

\caption{
    Hybrid one-qubit two-qumode circuit followed by measurements for computation of expectation values of a diagonal Hamiltonian as defined in \Eq{\ref{eq: diag_ham_qubit}}.
    The circuit consists of echoed conditional displacement (ECD) qubit-qumode gates with one-qubit rotations, as discussed in \Sec{\ref{sec: vqe}}.
}
\label{fig: full_circuit}
\end{figure}


The parameterized circuit for the trial state defined in \Eq{\ref{eq: ansatz}} must be a universal ansatz for the one qubit and multiple qumodes, which can be achieved in multiple ways. 
\cite{Krastanov2015,Liu2021sideband,Diringer2024cnot,You2024Crosstalk,Zhang2024energy}
We explore a universal ansatz based on the following circuit here \cite{You2024Crosstalk}
\begin{equation} \label{eq: full_uni_ecd_rot}
U (\mathbf{v})
= \UniER (\bm{\beta}_{N_d}, \bm{\theta}_{N_d}, \bm{\phi}_{N_d}) \cdots
\UniER (\bm{\beta}_1, \bm{\theta}_1, \bm{\phi}_1).
\end{equation}
The building block unitaries $U_{ER}$ are built from one-qubit arbitrary rotations 
\begin{equation}
R (\theta, \phi)
= e^{ - i (\theta / 2) \big[ \cos(\phi) X + \sin(\phi) Y \big] },   
\end{equation}
and two one-qubit one-qumode echoed conditional displacement (ECD) operations \cite{Eickbusch2022}
\begin{subequations} \label{eq: single_ecd_rot}
\begin{align} 
\UniER (\bm{\beta}_j, \bm{\theta}_j, \bm{\phi}_j)
&= ECD_{0, 2} (\beta_{j, 2}) \: 
R_0 (\theta_{j, 2}, \phi_{j, 2}) \nonumber
\\
&\times ECD_{0, 1} (\beta_{j, 1}) \:
R_0 (\theta_{j, 1}, \phi_{j, 1}),
\\
ECD_{0, 1} (\beta_1)
&= \sigma_0^- D_1 (\beta_1 / 2)
+ \sigma_0^+ D_1 (-\beta_1 / 2),
\\
ECD_{0, 2} (\beta_2)
&= \sigma_0^- D_2 (\beta_2 / 2)
+ \sigma_0^+ D_2 (-\beta_2 / 2),
\end{align}
\end{subequations}
where $ D (\beta) = e^{ \beta \BC{} - \beta^* \BA{} } $ is the qumode displacement operator, $X, Y$ are Pauli matrices, and 
$ \sigma^{+} = \KetBra{0}{1}, \sigma^{-} = \KetBra{1}{0} $ are the qubit transition operators. 
The operator subscripts in \Eq{\ref{eq: single_ecd_rot}} represent the indexing for the qubit and the two qumodes, and tensor product is assumed. 
Other choices for universal ansatz include selective number-dependent arbitrary phase (SNAP) with displacement and beamsplitters, \cite{Krastanov2015,liu2026hybrid} and conditional-not displacement gates. \cite{Diringer2024cnot}
The variables $\{ \bm{\beta}, \bm{\theta}, \bm{\phi} \}$ in \Eq{\ref{eq: full_uni_ecd_rot}} are matrices of dimensions $N_d \times 2$, where the complex-valued $\bm{\beta}$ matrix can also be split into two real-valued matrices of same dimensions such that 
$ \beta_{j, k} = r_{j, k} \: e^{i \tilde{\phi}_{j, k}} $ for each complex-valued element.
We call the number of blocks $N_d$ in \Eq{\ref{eq: full_uni_ecd_rot}} as the depth of the universal ECD-rotation circuit. 
Thus, the packed vector \textbf{v} representing all the real-valued parameters has $8 N_d$ dimensions.
The full circuit is illustrated in \Fig{\ref{fig: full_circuit}}.


\section{Model Applications} \label{sec: application}

We now have all the tools needed to implement the ECD-VQE method for finding the ground state of a diagonal Hamiltonian. 
The examples below are intended to illustrate three aspects of the method: 
(i) how a constrained optimization problem is mapped to a diagonal qubit--qumode Hamiltonian, 
(ii) how the optimal bitstring is identified from the joint qubit and photon-number measurement distribution, and 
(iii) how the hybrid implementation compares with qubit-only alternatives in circuit depth and resource requirements. 
Because the goal is combinatorial optimization rather than high-precision energy estimation, we focus not only on convergence of the variational energy but also on the probability assigned to the optimal basis state.

\subsection{Binary knapsack problem} \label{sec: bkp_example}


\begin{figure}[t!]

\includegraphics[width=0.9\columnwidth]{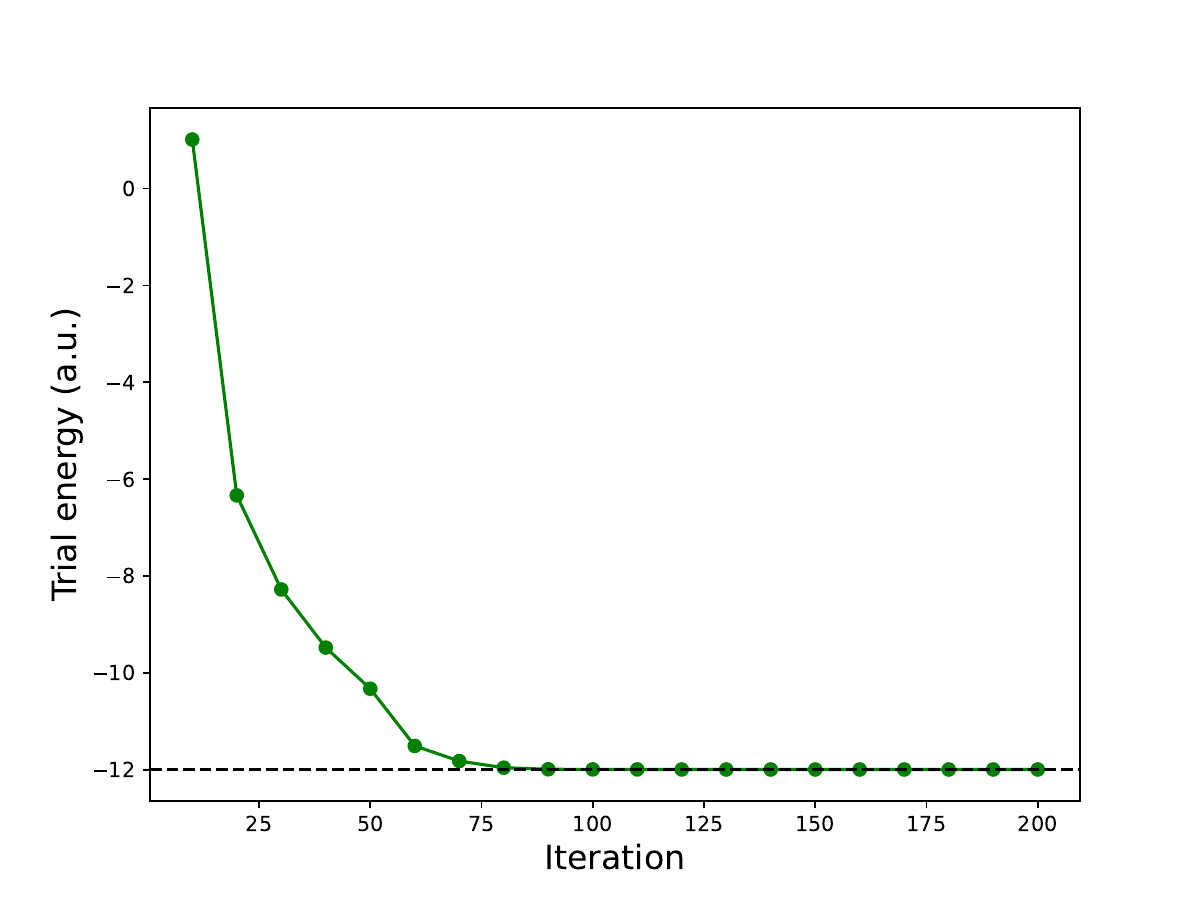}

\caption{
    Trial energy values defined in \Eq{\ref{eq: vqe}} at different ECD-VQE iterations while finding the ground state of $H_Q$ defined in \Eq{\ref{eq: bkp_q7_example_ham}}. 
    The horizontal black line represents the exact ground state energy.
    The circuit depth for the trial state is $N_d = 5$.
}
\label{fig: bkp_energies_lval2_nd5}
\end{figure}

Let us discuss a simple binary knapsack problem.
Let us assume we have $N_0 = 4$ items with their values and weight constraints given by 
\begin{subequations} \label{eq: bkp_q7_example}
\begin{align}
\max_{\mathbf{x}} \quad 
&2 \: x_0 + 5 \: x_1
+ 7 \: x_2 + 3 \: x_3,
\\
\text{subject to} \quad 
&2.5 \: x_0 + 3 \: x_1
+ 4 \: x_2 + 3.5 \: x_3 \leq 7.
\end{align}
\end{subequations}
Following the discussions in \Sec{\ref{sec: bkp_qubo_qubit}}, the above optimization can be recast as the following QUBO problem 
\begin{align} 
\min_{\mathbf{x}} E 
&= - ( 2 \: x_0 + 5 \: x_1
+ 7 \: x_2+ 3 \: x_3 ) \nonumber
\\
&+ \lambda \: \Big[ 7
- ( 2.5 \: x_0 + 3 \: x_1
+ 4 \: x_2 + 3.5 \: x_3 ) \nonumber
\\
&\hskip3ex
- ( x_4 + 2 \: x_5 + 4 \: x_6 ) \Big]^2
\end{align}
consisting of 4 + 3 = 7 binary variables. 
For the penalty weight $\lambda = 2$, the corresponding seven-qubit Hamiltonian is given by 
\begin{align} \label{eq: bkp_q7_example_ham}
H_Q 
&= 41.75 
- 14.0 \: Z_0
- 15.5 \: Z_1
- 20.5 \: Z_2
- 19.5 \: Z_3
\nonumber
\\
&- 6.0 \: Z_4
- 12.0 \: Z_5
- 24.0 \: Z_6
+ 7.5 \: Z_0 Z_1 
+ 10.0 \: Z_0 Z_2
\nonumber
\\
&+ 8.75 \: Z_0 Z_3 
+ 2.5 \: Z_0 Z_4
+ 5.0 \: Z_0 Z_5
+ 10.0 \: Z_0 Z_6
\nonumber
\\
&+ 12.0 \: Z_1 Z_2
+ 10.5 \: Z_1 Z_3
+ 3.0 \: Z_1 Z_4
+ 6.0 \: Z_1 Z_5
\nonumber
\\
&+ 12.0 \: Z_1 Z_6
+ 14.0 \: Z_2 Z_3 
+ 4.0 \: Z_2 Z_4
+ 8.0 \: Z_2 Z_5
\nonumber
\\
&+ 16.0 \: Z_2 Z_6
+ 3.5 \: Z_3 Z_4
+ 7.0 \: Z_3 Z_5
+ 14.0 \: Z_3 Z_6
\nonumber
\\
&+ 2.0 \: Z_4 Z_5
+ 4.0 \: Z_4 Z_6
+ 8.0 \: Z_5 Z_6.
\end{align}


\begin{figure*}[t!]

\includegraphics[width=0.9\textwidth]{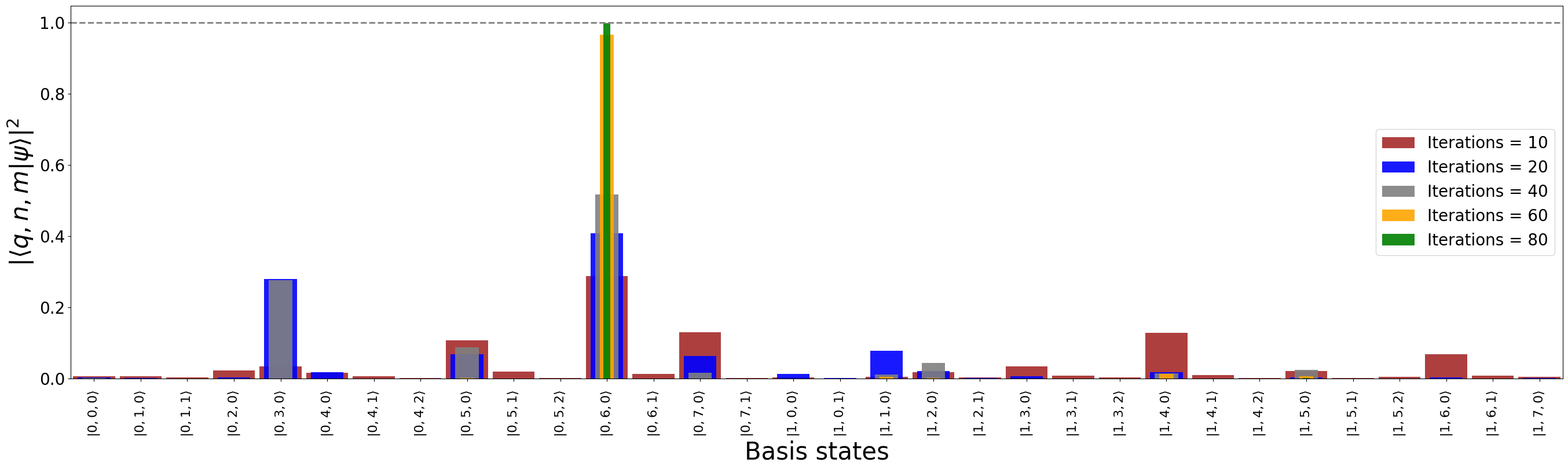}

\caption{
    Probabilities 
    $ S_{q, n, m} = | \braket{q, n, m | \psi} |^2 $ at different numbers of iterations of the ECD-VQE method for the one-qubit two-qumode Hamiltonian $H_Q$ defined in \Eq{\ref{eq: bkp_q7_example_ham}}.
    The histograms are split into two parts for better readability of the basis states.  
    The circuit depth for the trial state is $N_d = 5$.
    The corresponding trial energy values are shown in \Fig{\ref{fig: bkp_energies_lval2_nd5}}. For readability, only basis states with measurement probabilities above the plotting threshold are shown; all basis states were retained in the simulations and optimization.
}
\label{fig: bkp_probs_nd5_niters}
\end{figure*}


{\bf Qumode-based ECD-VQE:} The ground state of $H_Q$ obtained by exact diagonalization is 
$ \ket{0, 1, 1, 0}_Q \otimes \ket{0, 0, 0}_Q $ with eigenvalue -12.
Indeed, the optimal solution is 
$\mathbf{x}^* = (0, 1, 1, 0)$ with the corresponding weight = 7 and value = 12. 
We will now partition the seven-qubit Hamiltonian into three parts such that it will be mapped to a one-qubit two-qumode system with the Fock cutoff for each qumode being $L = 8$. 
In other words, we map the first four qubits corresponding to the primary variables to the qubit along with the first qumode, and the rest to the second qumode. 
The ground state of $H_Q$ is now mapped as
\begin{equation}
\ket{0, 1, 1, 0}_Q \otimes \ket{0, 0, 0}_Q 
\leftrightarrow \ket{0}_Q \otimes \ket{6}_B \otimes \ket{0}_B, 
\end{equation}
or in shorthand, $\ket{0, 6, 0}$.
This is the solution state that we seek with this model benchmark system.

We show the trial energies for the ECD-VQE method applied to $H_Q$ of \Eq{\ref{eq: bkp_q7_example_ham}} in \Fig{\ref{fig: bkp_energies_lval2_nd5}}.
The circuit depth chosen for the trial state is $N_d = 5$.
We emulated the expectation values classically using the QuTiP Python library \cite{Johansson2012qutip} and optimized the energy function of \Eq{\ref{eq: vqe}} using the Broyden--Fletcher--Goldfarb--Shanno (BFGS) algorithm as implemented in the SciPy Python library. \cite{Virtanen2020scipy}
It is clear from \Fig{\ref{fig: bkp_energies_lval2_nd5}} that the ECD-VQE method discussed here practically converges to the exact ground state energy in approximately 100 iterations. 
The goal in traditional VQE approaches is usually to find highly accurate ground state energies, whereas our goal here is to resolve the ground state $\ket{0, 6, 0}$, which represents the optimal solution.
We plot the corresponding $\{ S_{q, n, m} \}$ probability values as defined in \Eq{\ref{eq: probs}} during different iterations of ECD-VQE in \Fig{\ref{fig: bkp_probs_nd5_niters}}.
The ground state is practically resolved after 80 iterations and emerges as the highest peak even in 10 iterations, as shown in \Fig{\ref{fig: bkp_probs_nd5_niters}}.
Thus, resolving the ground state may be achieved using a relatively smaller number of classical optimization steps than for finding the energy.
We highlight that the $\{ S_{q, n, m} \}$ probability values are directly available from photon number and Pauli-$Z$ measurement histograms without the need for explicitly iterating over all possible integer strings, such as in a classical simulator. 

{\bf Qubit-based ECD-VQE:} Implementing the ECD-VQE ansatz on a qubit-only device would result in extremely deep circuits, \cite{Wang2020vibronic,liu2026hybrid} 
motivating our use of hybrid qubit-qumode hardware.
To estimate the equivalent qubit-only resource cost, we transpile the qubit-qumode unitaries using Qiskit, \cite{Qiskit2024} 
employing Pauli rotations and CNOT gates as the elementary operations. 
For this problem, with a Fock cutoff $L = 8$ per qumode, each qumode becomes encoded using four qubits, including an ancilla.
Transpilation shows that a single ECD gate maps to 93 CNOT gates and yields a circuit depth of approximately $270$ on a four-qubit register. 
Consequently, the ECD-rotation ansatz with $N_d = 5$ blocks, mentioned above, comprising 10 ECD gates, corresponds to a total of 930 CNOT gates in the qubit-only model. 
Alternatively, compiling the full ansatz (with $N_d = 5$) into a single unitary acting on $N = 7$ qubits results in $7660$ CNOT gates and a circuit depth of around $21400$. 
These results highlight that qubit-qumode implementations can achieve comparable expressivity with significantly reduced circuit depth relative to its qubit counterparts. 


\begin{figure}[h!]
    \centering
    \includegraphics[width=0.9\columnwidth]{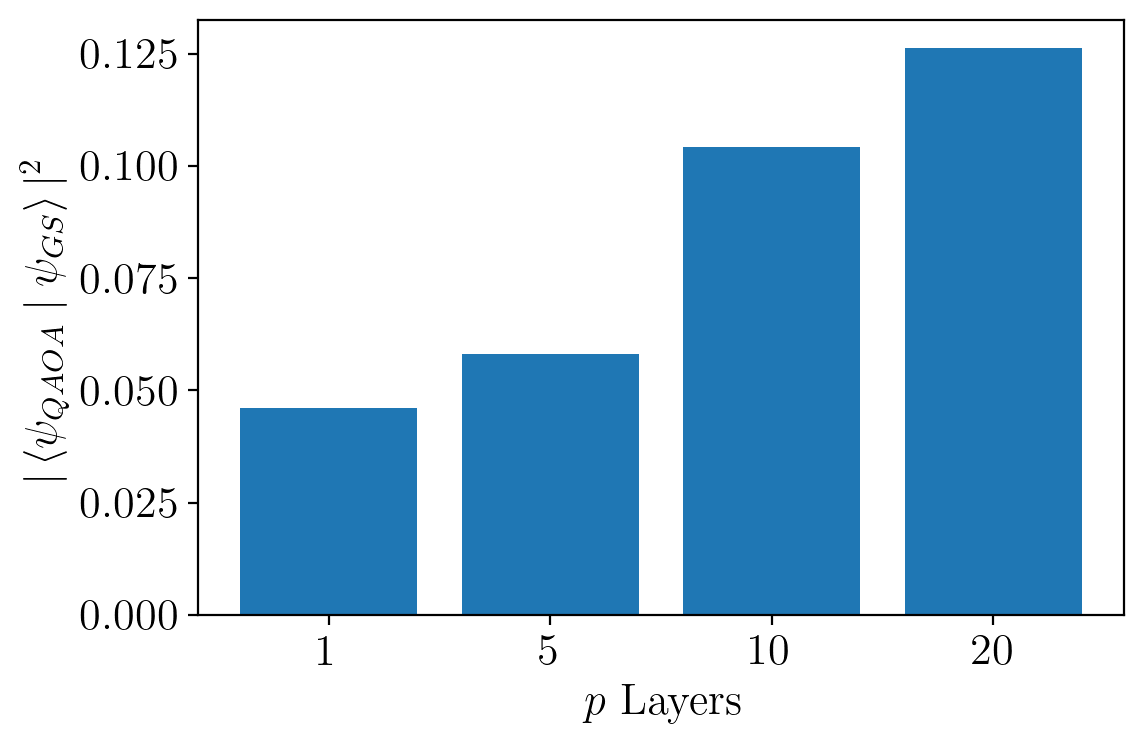}
    \caption{
        Optimal probabilities for different numbers of QAOA layers chosen out of 50 independent trials for the ground state of the seven-qubit Hamiltonian defined in  \Eq{\ref{eq: bkp_q7_example_ham}}. 
        Each trial converged with $\sim 150$ iterations with the classical optimizer. 
    }
    \label{fig:bkp_qaoa_best_probs_nlayers}
\end{figure}


\begin{figure*}[t!]
    \centering
    \includegraphics[width=0.9\textwidth]{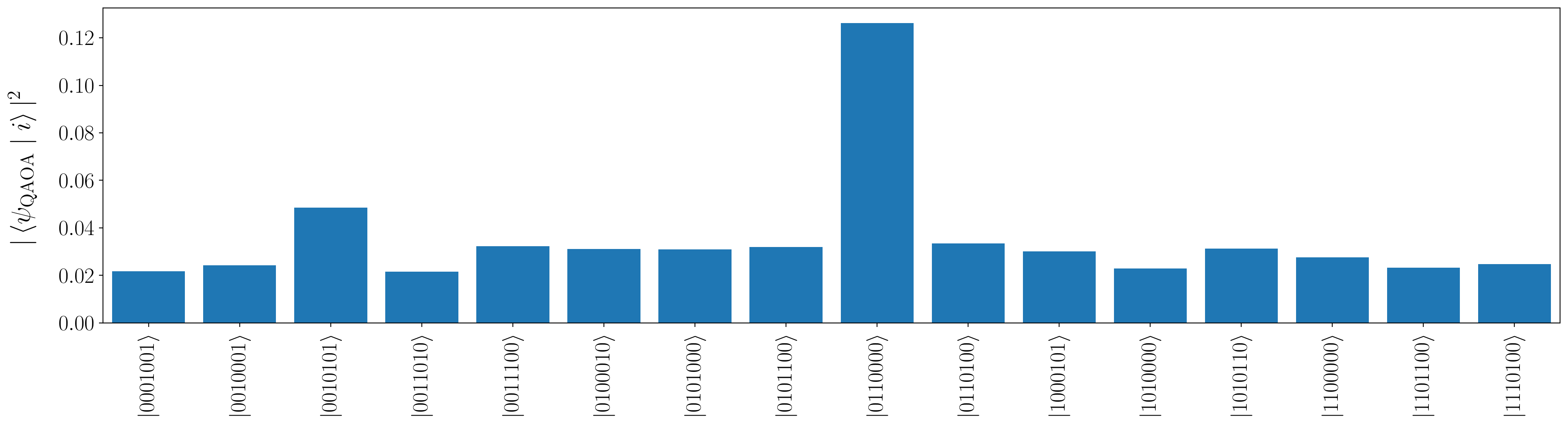}
    \caption{
         Measurement probabilities for bitstring basis states for the optimal QAOA state with $p=20$-layers for the seven-qubit Hamiltonian defined in \Eq{\ref{eq: bkp_q7_example_ham}}. For readability, only basis states with measurement probabilities above the plotting threshold are shown; all basis states were retained in the simulations and optimization.
    }
    \label{fig:bkp_qaoa_20layers_samples}
\end{figure*}


{\bf Qubit-based QAOA:} Here, we compare our optimization results for the BKP problem to those obtained using the Quantum Approximate Optimization Algorithm (QAOA) implemented with standard qubit-based circuit \cite{Farhi2014quantum,Blekos2024review}. 
The QAOA implementation is a qubit-only VQA commonly used as a benchmark for combinatorial binary optimization problems.
QAOA can be thought of as a special case of VQE, where the PQC ansatz consists of the alternating application of $p$-layers of parametrized ``mixing" and ``problem" unitaries, applied to a quantum register initialized in a uniform superposition
\begin{align}
|\psi(\beta, \gamma)\rangle 
&= e^{-i \beta_p H_M} e^{-i \gamma_p H_P} \: \cdots \: 
e^{-i \beta_2 H_M} e^{-i \gamma_2 H_P} \nonumber
\\
&\hskip3ex e^{-i \beta_1 H_M} e^{-i \gamma_1 H_P} \ket{+}^{\otimes N}.
\end{align}
The problem Hamiltonian $H_P$ in this cased is same as in \Eq{\ref{eq: bkp_q7_example_ham}}, whereas the mixing Hamiltonian is defined as 
\begin{equation}
H_M 
= \sum_{j=0}^{N-1} \sigma_{j}^{x}.
\end{equation}
The QAOA calculations were implemented numerically with QuTiP using the BFGS classical optimizer. 

The optimal QAOA result out of 50 independent trials for increasing layers is shown in \Fig{\ref{fig:bkp_qaoa_best_probs_nlayers}}. 
These results show that the probability of measuring the bitstring corresponding to the solution generally increases by increasing the number of layers.  
We note that the QAOA approach with $p = 20$ layers has the same number of variational parameters as the ECD-VQE approach with $N_d = 5$ blocks, highlighting the more favorable optimization landscape for the ECD-based approach.  
We also note that the number of CNOT gates per layer scales as $\ComCom{N^2}$ for QUBO problems where $N$ is the number of qubits \cite{Weidenfeller2022}. 
Indeed, the number of two-qubit CNOT gates needed for the BKP problem discussed here for only one layer is 42, which is more than the gate counts for the ECD-VQE approach, where $N_d = 5$ blocks correspond to only 10 one-qubit one-qumode ECD gates. 

Since the highest probability of measuring the solution was obtained with $p=20$ layers, we chose to sample the measurement outcomes with this optimal QAOA circuit. 
The results of the measurement sampling are shown in \Fig{\ref{fig:bkp_qaoa_20layers_samples}}. 
Although the solution bitstring has the highest measurement probability, we see that there is still a substantial likelihood of sampling sub-optimal states. 
The comparison between \Fig{\ref{fig: bkp_probs_nd5_niters}} and \Fig{\ref{fig:bkp_qaoa_20layers_samples}} indicates the potential advantages of an expressive VQE ansatz over the QAOA approach, whose ansatz is limited by its cost and mixing Hamiltonians. 
In our case, the VQE ansatz is provided by a set of native qubit-qumode gates that reveal the optimal solutions with the help of a few blocks of gates. 

Although the BKP instance studied here is intentionally small, the same mapping applies directly to larger knapsack problems. 
For a problem with $N_0$ items and capacity $W$, the QUBO construction requires 
$N_{\rm BKP}=N_0+\lceil\log_2(W+1)\rceil$ binary variables, including the auxiliary variables used to encode the capacity constraint. 
Such an instance can be embedded on a hybrid device whenever 
$N_{\rm BKP}\leq N_{\rm bin}^{\max}$, with $N_{\rm bin}^{\max}$ determined by the available qubits, qumodes, and Fock cutoffs as discussed in Sec.~\ref{sec: final}. 
Thus, scaling to larger BKP instances is achieved by adding qumodes and/or qubits rather than by assuming arbitrarily large photon-number cutoffs. 
The resulting Hamiltonian remains diagonal but contains $\mathcal{O}(N_{\rm BKP}^2)$ coefficients, so large-scale BKP applications remain heuristic optimization problems and require further benchmarking.

\subsection{Multiple constraints}


\begin{figure}[t!]

\includegraphics[width=0.9\columnwidth]{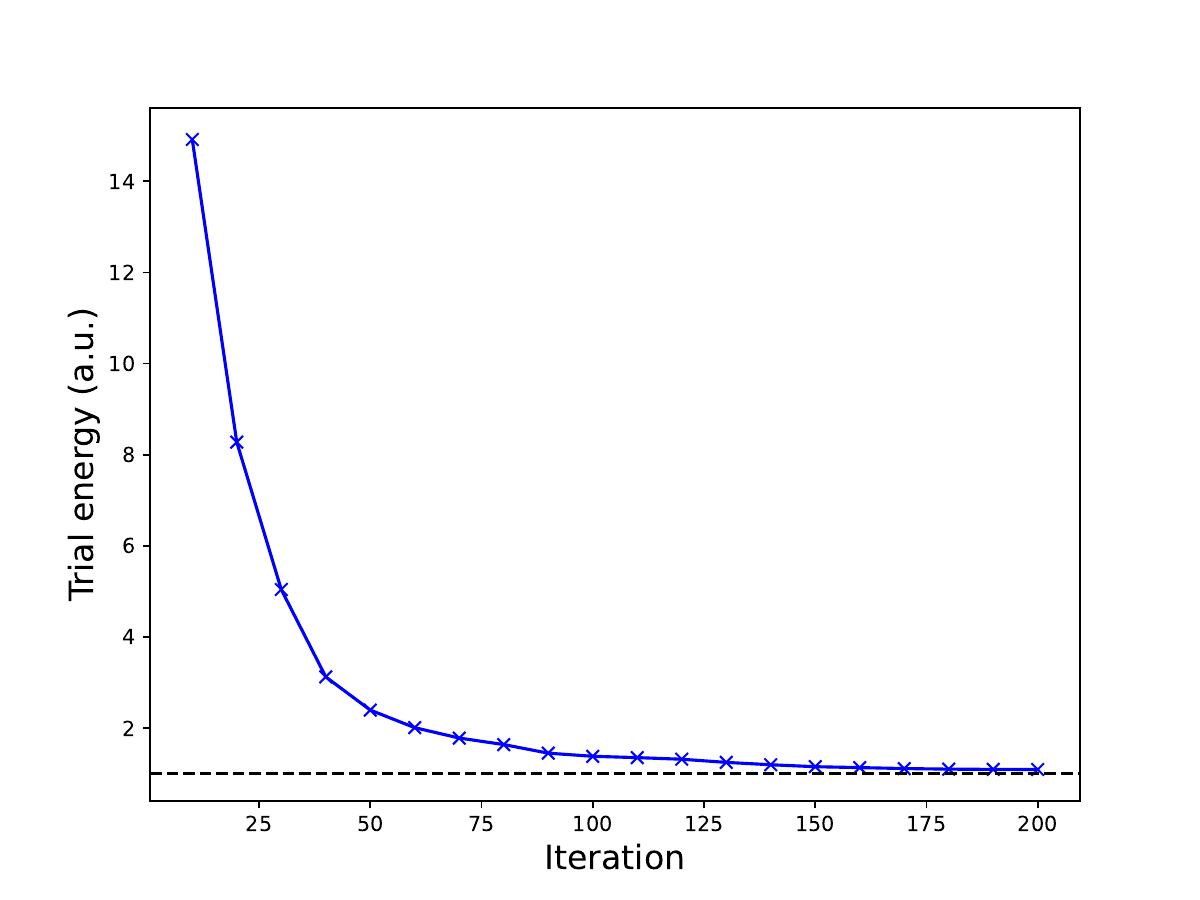}

\caption{
    Trial energy values defined in \Eq{\ref{eq: vqe}} at different ECD-VQE iterations while finding the ground state of $H_Q$ defined in \Eq{\ref{eq: mc_q6_example_ham}}. 
    The horizontal black line represents the exact ground state energy.
    The circuit depth for the trial state is $N_d = 10$.
}
\label{fig: mc_energies_lval5_nd10}
\end{figure}

Our approach can be applied to any constrained optimization problem beyond the BKP problem discussed above. 
Indeed, let us discuss another simple constrained optimization problem given below 
\begin{subequations} \label{eq: mc_q6_example}
\begin{align}
\min_{\mathbf{x}} \quad &
x_0 + 2 x_1 + x_2,
\\
\text{subject to} \quad &
x_0 + x_1 = 1,
\\
&2 x_0 + 2 x_1 + x_2 \leq 3,
\\
&x_0 + x_1 + x_2 \geq 1.
\end{align}
\end{subequations}
which can be represented as the following QUBO problem 
\begin{align} 
\min_{\mathbf{x}} F
&= x_0 + x_1 + x_2
+ \lambda_1 \: ( 1 - x_0 - x_1 )^2
\nonumber
\\
&+ \lambda_2 \: \big[ 3 - ( 2 x_0 + 2 x_1 + x_2 )
- ( x_3 + 2 x_4 ) \big]^2
\nonumber
\\
&+ \lambda_3 \: \big[ ( x_0 + x_1 + x_2 )
- x_5 - 1 \big]^2,
\end{align}
consisting of 3 + 3 = 6 binary variables. 
For the penalty weights $\lambda_1 = \lambda_2 = \lambda_3 = 5$, the corresponding six-qubit Hamiltonian is given by 
\begin{align} \label{eq: mc_q6_example_ham}
H_Q 
&= 32.0
- 10.5 \: Z_0
- 11.0 \: Z_1 
- 5.5 \: Z_2
- 5.0 \: Z_3 \nonumber
\\
&- 10.0 \: Z_4 
+ 15.0 \: Z_0 Z_1 
+ 7.5 \: Z_0 Z_2
+ 5.0 \: Z_0 Z_3 \nonumber
\\
&+ 10.0 \: Z_0 Z_4 
- 2.5 \: Z_0 Z_5
+ 7.5 \: Z_1 Z_2
+ 5.0 \: Z_1 Z_3 \nonumber
\\
&+ 10.0 \: Z_1 Z_4 
- 2.5 \: Z_1 Z_5
+ 2.5 \: Z_2 Z_3
+ 5.0 \: Z_2 Z_4 \nonumber
\\
&- 2.5 \: Z_2 Z_5
+ 5.0 \: Z_3 Z_4.
\end{align}
The ground state of $H_Q$ obtained by exact diagonalization is 
$ \ket{1, 0, 0}_Q \otimes \ket{1, 0, 0}_Q $ with eigenvalue 1.
Indeed, the optimal solution is 
$\mathbf{x}^* = (1, 0, 0)$.
Let us now reorganize the Hilbert space of the six-qubit Hamiltonian into three parts such that it will be mapped to a one-qubit two-qumode system with the Fock cutoffs for the two qumodes being $L_1 = 4$ and $L_2 = 8$. 
In other words, we map the first three qubits corresponding to the primary variables to the qubit along with the first qumode, and the rest to the second qumode. 
The ground state of $H_Q$ is now mapped as
\begin{equation}
\ket{1, 0, 0}_Q \otimes \ket{1, 0, 0}_Q 
\leftrightarrow \ket{1}_Q \otimes \ket{0}_B \otimes \ket{4}_B, 
\end{equation}
or in shorthand, $\ket{1, 0, 4}$ will be our target state for the ECD-VQE approach as before.


\begin{figure*}[t!]

\includegraphics[width=0.95\textwidth]{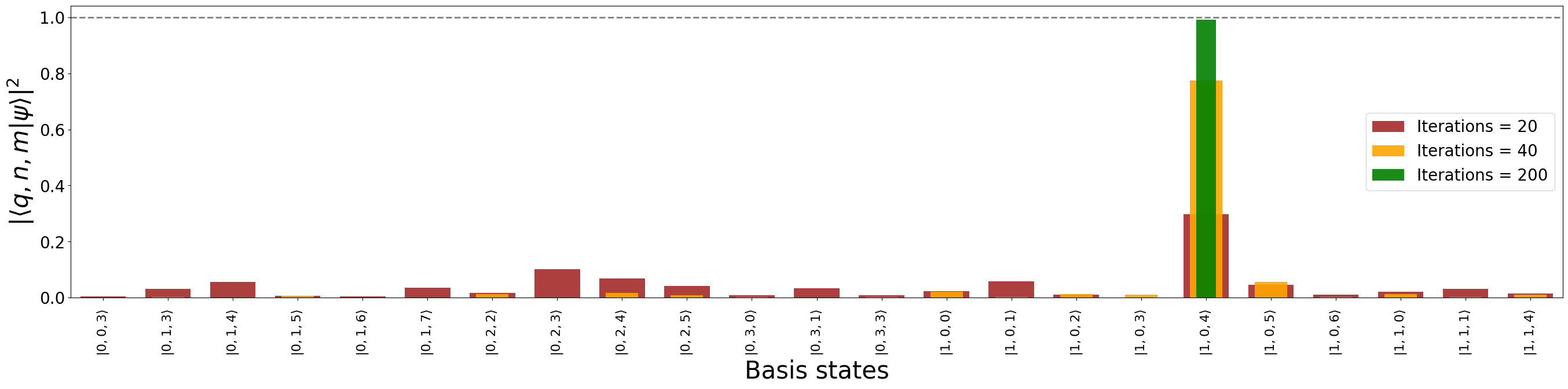}

\caption{
    Probabilities 
    $ S_{q, n, m} = | \braket{q, n, m | \psi} |^2 $ at different numbers of iterations of the ECD-VQE method for the one-qubit two-qumode Hamiltonian $H_Q$ defined in \Eq{\ref{eq: mc_q6_example_ham}}.
    The circuit depth for the trial state is $N_d = 10$.
    The corresponding trial energy values are shown in \Fig{\ref{fig: mc_energies_lval5_nd10}}. For readability, only basis states with measurement probabilities above the plotting threshold are shown; all basis states were retained in the simulations and optimization.
}
\label{fig: mc_probs_nd10_niters}
\end{figure*}


We show the trial energies for the ECD-VQE method applied to $H_B$ of \Eq{\ref{eq: mc_q6_example_ham}} in \Fig{\ref{fig: mc_energies_lval5_nd10}} and the corresponding overlaps in \Fig{\ref{fig: mc_probs_nd10_niters}}.
The circuit depth chosen for the trial state is $N_d = 10$.
It is clear that the ground state of the Hamiltonian is fairly resolved after 40 iterations and emerges as the highest peak even in 20 iterations, as shown in \Fig{\ref{fig: mc_probs_nd10_niters}}, even though there is space for the trial energy to converge to lower values even after 200 iterations.
This again shows the efficiency of this method in resolving the optimal solution state with just a few optimization iterations.


\section{Active space selection for multireference methods}
\label{sec: chem_qubo}

Many computational chemistry tasks can be expressed as discrete optimization problems with hard constraints, including configuration selection, determinant pruning, and active-space construction. \cite{ma2011generalized, schriber2016communication, stein2016automated, huron1973iterative, olsen1988determinant}  These problems often require choosing a subset of orbitals or configurations that satisfy fixed-size or symmetry constraints while maximizing some measure of correlation or importance.  Such combinatorial structure makes them natural candidates for formulation as quadratic unconstrained binary optimization (QUBO) problems, which can be implemented directly as diagonal Hamiltonians on the hybrid qubit--qumode architecture developed in the preceding sections.  

In this section, we focus on one representative example: the construction of active spaces for multireference methods such as CASCI and CASSCF.\cite{roos1980complete}  Selecting an appropriate subset of molecular orbitals is a longstanding challenge, as some orbitals exhibit negligible correlation while others form strongly entangled groups that must be included together.  We show that active-space selection admits a compact and physically interpretable QUBO formulation, enabling its solution via the ECD-VQE framework introduced earlier.

\subsection{Mapping to a weighted graph}

For a molecule with $n$ spatial orbitals, we assign to each orbital $i \in \{1,\dots,n\}$ two types of correlation indicators:

\begin{enumerate}
\item \emph{Single-orbital entropies}\cite{ActiveSpaceFinder} $s_i \ge 0$, computed from the one-orbital reduced density matrices, which quantify how entangled orbital $i$ is with the rest of the system.
\item \emph{Pairwise correlation weights}\cite{ActiveSpaceFinder} $I_{ij} = \kappa_{ij,ij} \ge 0$, derived from the diagonal elements of the two-body cumulant
\begin{equation}
\kappa_{pq,rs}
=
\Gamma_{pq,rs}
-
\big(\gamma_{pr}\gamma_{qs} - \gamma_{ps}\gamma_{qr}\big),
\end{equation}
where $\gamma$ and $\Gamma$ denote the one- and two-body reduced density matrices, respectively.
\end{enumerate}

Together, $\{s_i\}$ and $\{I_{ij}\}$ define a weighted graph: nodes correspond to orbitals and carry weights $s_i$, while edges encode pairwise correlation via $I_{ij}$.  
Intuitively, strongly entangled or strongly coupled orbitals should be selected together in an active space.

\subsection{QUBO formulation}

Let $x_i \in \{0,1\}$ indicate whether orbital $i$ is included in the active space.  
For a target active-space size $K$, we seek the set of $K$ orbitals that maximizes the total correlation captured by the chosen subset.  
This produces a QUBO objective with soft and hard constraints,
\begin{align}
\max_{x \in \{0,1\}^n} \
& \alpha \sum_i s_i x_i
+ \beta \sum_{i<j} I_{ij} x_i x_j, \\
\text{s.t.}\quad &
\sum_i x_i = K , \nonumber
\end{align}
which we convert to an unconstrained energy function,
\begin{align} \label{eq:chem_qubo}
E_{\text{QUBO}}(x)
&= - \alpha \: \sum_i s_i x_i
- \beta \: \sum_{i<j} I_{ij} x_i x_j
\nonumber
\\
&+ \lambda_K \bigg(\sum_i x_i - K\bigg)^2.
\end{align}
Here, $\alpha$ and $\beta$ balance single-orbital and pairwise correlation metrics, while $\lambda_K$ enforces the fixed-size constraint.  
This form matches the general QUBO structure used in Secs.~\ref{sec: mapping}–\ref{sec: vqe}, enabling direct mapping to a diagonal Hamiltonian.

To embed the problem onto one qubit and two qumodes, we partition the Hilbert space such that
\[
2^N = 2 \times L_1 \times L_2,
\]
with Fock cutoffs $L_1$ and $L_2$ chosen to accommodate the $N$ binary variables.  
The resulting diagonal Hamiltonian is then evaluated variationally using the ECD-VQE ansatz, and the optimal bitstring is recovered from the qubit $Z$ measurement and qumode photon-number readouts.

\begin{figure*}[t!]

\includegraphics[width=0.95\textwidth]{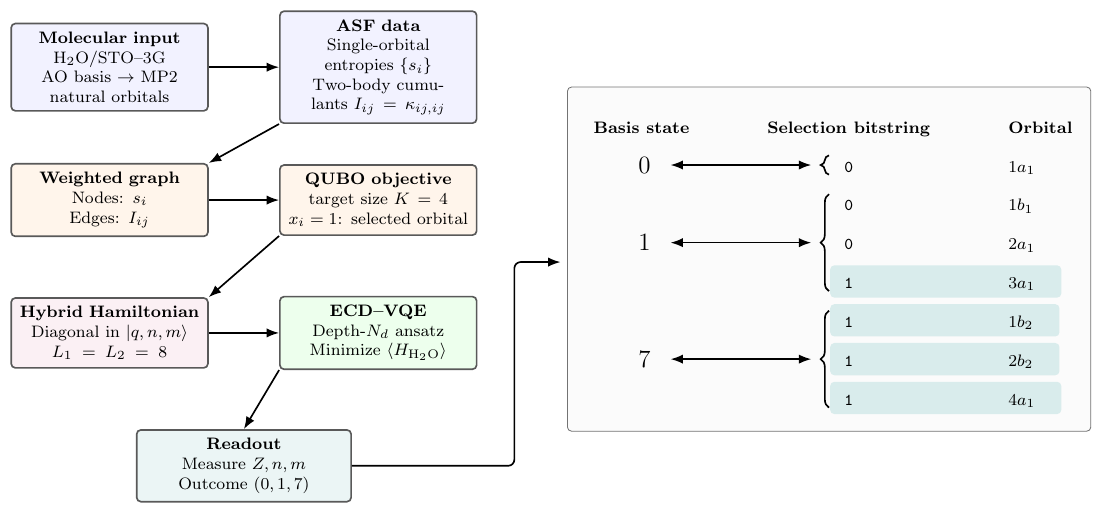}

\caption{
End-to-end workflow for the water active-space-selection example.
Starting from the $\mathrm{H_2O}$/STO--3G atomic-orbital description, the ASF workflow transforms to a molecular-orbital basis and computes single-orbital entropies $s_i$ and pairwise cumulant weights $I_{ij}=\kappa_{ij,ij}$.
These quantities define a weighted graph and the fixed-size QUBO objective in Eq.~\eqref{eq:chem_qubo} with $K=4$.
The resulting diagonal Hamiltonian is embedded in one qubit and two qumodes with $L_1=L_2=8$, so $2^7=2\times8\times8$.
ECD--VQE is used to approximate the ground state of the hybrid Hamiltonian.
The optimized measurement outcome $(q,n,m)=(0,1,7)$ maps to $x=0\,|\,001\,|\,111$, selecting the highlighted orbitals $3a_1$, $1b_2$, $2b_2$, and $4a_1$.
}
\label{fig:bit-to-basis}
\end{figure*} 

\subsection{Case study: Water Molecule}

As a benchmark demonstration, we consider the active space selection for the water molecule in the STO-3G basis from the equilibrium configuration which contains $n=7$ spatial orbitals: five oxygen orbitals ($1s$, $2s$, $2p_x$, $2p_y$, $2p_z$) and two hydrogen $1s$ orbitals, ASF then transforms this AO basis to a MO basis of MP2 natural orbitals, yielding 7 orthonormal spatial MOs.  
Using the Active Space Finder (ASF) workflow,~\cite{ActiveSpaceFinder} we compute the single-orbital entropies $\{s_i\}$ and the diagonal two-body cumulant elements $\{I_{ij}\}$ in a consistent molecular-orbital basis.

We construct a QUBO instance from Eq.~\eqref{eq:chem_qubo} with a target active-space size of $K=4$, corresponding to a compact valence description.  
The resulting seven-variable QUBO is then mapped to a qubit--qumode Hamiltonian using Fock cutoffs $L_1=L_2=8$, giving
\[
2^7 = 2 \times 8 \times 8,
\]
so that the hybrid Hilbert space matches the dimensionality of the equivalent seven-qubit Ising problem. The full workflow, from ASF correlation metrics to QUBO construction, hybrid qubit--qumode embedding, ECD--VQE solution, and final orbital-selection decoding, is illustrated in Figure~\ref{fig:bit-to-basis}.
The diagonal Hamiltonian $H_Q$ is encoded following Sec.~\ref{sec: mapping}, and its ground state is approximated with the ECD-VQE procedure from Sec.~\ref{sec: vqe}. 

Photon-number measurements on the two qumodes, combined with the qubit measurement, uniquely label the basis states $\ket{q,n,m}$ and identify the optimal active-space bitstring.

\begin{figure}[h!]

\includegraphics[width=0.25\textwidth]{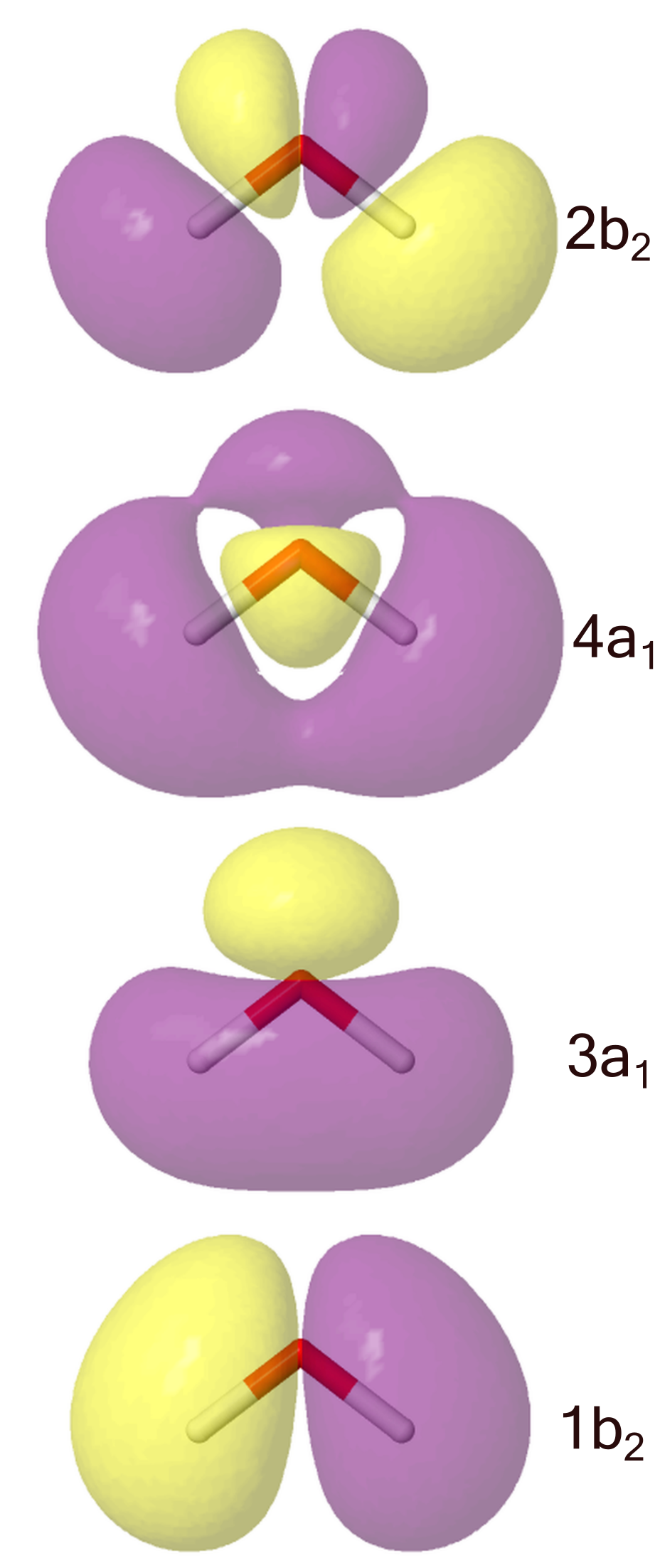}

\caption{
Visual representations of the H$_2$O molecular orbitals comprising the selected active space of size 4 with colors indicating the orbital phase.
}
\label{fig: water_selected_K4}
\end{figure}

In our numerical experiments, the ECD‑VQE optimum matches the ASF results, which corresponds to the MO set
$\{3a_1,\,1b_2,\,2b_2,\,4a_1\}$: one oxygen lone pair ($3a_1$), the $b_2$
$\sigma/\sigma\mbox{*}$ pair ($1b_2/2b_2$), and the $a_1$ $\sigma\mbox{*}$ ($4a_1$).
The deeply bound core orbital ($1a_1$) and the remaining lone pair ($1b_1$)
are excluded. This compact choice captures the dominant static correlation at
the equilibrium geometry while keeping the active space minimal. Real-space visualizations of these selected molecular orbitals are shown in Figure~\ref{fig: water_selected_K4}.

\begin{figure*}[ht]

\includegraphics[width=0.95\textwidth]{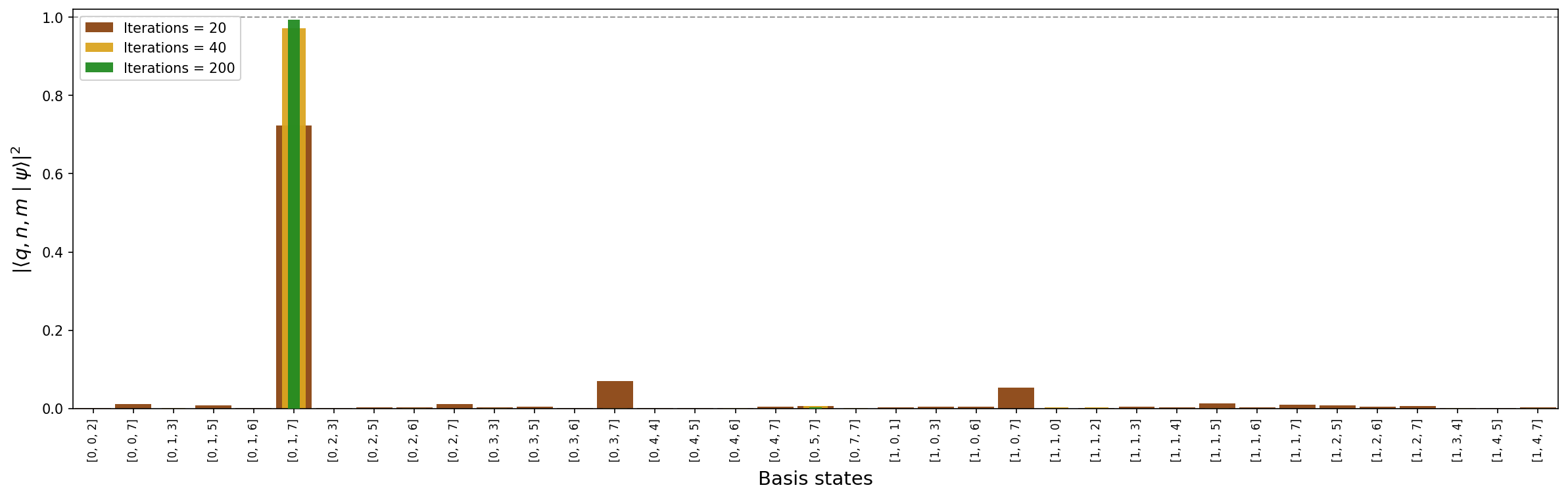}

\caption{
        Final probability distribution $|\braket{q,n,m|\psi}|^2$ for the ECD-VQE ground-state approximation of the H$_2$O active-space QUBO.
        The dominant peak corresponds to the optimal bitstring, which closely agrees with the ASF-selected active space. The circuit depth for the trial state is $N_d = 10$. For readability, only basis states with measurement probabilities above the plotting threshold are shown; all basis states were retained in the simulations and optimization.
}
\label{fig: h2o_probs_nd10_niters}
\end{figure*}

Figure~\ref{fig: h2o_probs_nd10_niters} displays the final probability distribution over the hybrid basis states.  
The peak associated with the correct bitstring dominates the distribution even at moderate circuit depth and before full convergence of the variational energy.  
This mirrors the performance observed in the knapsack and multi-constraint benchmarks of Sec.~\ref{sec: application}, and suggests that active-space identification can be performed with shallow ECD circuits and a modest number of classical optimization steps.

The water example should be viewed as a proof-of-principle demonstration of the full active-space-selection workflow rather than a comprehensive benchmark of active-space selection across chemical space. 
The same QUBO construction can be applied to other molecules whenever the corresponding orbital entropies, cumulant-based pair weights, and target active-space size are available, but the quality of the resulting active space remains dependent on the chosen orbital basis, molecular geometry, correlation diagnostics, and hardware-accessible problem size. 
A systematic benchmark over larger molecules and basis sets is left for future work.

\section{Discussion} \label{sec: final}

We have demonstrated that ECD-VQE provides an effective and hardware-efficient strategy for solving optimization problems with soft and hard constraints on hybrid qubit-qumode quantum devices available in cQED architectures.
By leveraging only two qumodes coupled to a single transmon qubit---corresponding physically to two microwave cavities dispersively coupled to one nonlinear element---we showed that problems normally requiring several qubits can be embedded into a compact hybrid Hilbert space. 
Generalization to additional qumodes is straightforward, enabling scalable representations of larger QUBO instances. 

Our benchmarks included the binary knapsack problem as well as a multiple-constraint optimization task mapped from a six-qubit Hamiltonian, illustrating how multi-qubit diagonal operators can be compressed into a qubit--qumode representation. 
In Sec.~\ref{sec: chem_qubo}, we also extended this framework to chemically inspired optimization, demonstrating that active-space selection for multireference electronic structure methods can be expressed as a QUBO problem and efficiently solved using the same hybrid architecture.

The ECD-VQE workflow consists of preparing the qubit--qumode circuit, followed by joint measurements of the qubit state and the qumode photon numbers to evaluate expectation values of the mapped Hamiltonian. 
In many cases, these measurements identify the correct optimal configuration even before the variational energy has fully converged to the ground-state value. 
This early resolution of high-probability bitstrings may reduce the number of optimization iterations needed compared to traditional VQE approaches. \cite{Lee2023evaluating}
Although classical QUBO heuristics such as simulated annealing \cite{Suman2006survey} are important baselines, their performance depends strongly on the problem structure and implementation details, making a systematic comparison beyond the scope of this proof-of-principle study. 
Here, we benchmark against a standard qubit-based QAOA implementation for the BKP example and against the Active Space Finder workflow as a chemistry-specific reference for active-space selection.

The Hilbert space mapping strategy of Sec.~\ref{sec: mapping} applies to any multi-qubit Hamiltonian composed solely of diagonal Pauli-$Z$ and identity operators, which includes all QUBO formulations. 
Thus, any QUBO can be translated into a diagonal qubit Hamiltonian and subsequently compressed into the tensor product space of one qubit and multiple qumodes. 
To make this scalability explicit, the maximum QUBO size that can be embedded is set by the total accessible hybrid Hilbert-space dimension rather than by the number of physical transmons and cavities alone. 
More generally, with $Q$ computational qubits and $R$ qumodes of Fock cutoffs $\{L_j\}$, the number of binary variables that can be represented satisfies
\[
2^{N_{\rm bin}} \leq 2^Q \prod_{j=1}^R L_j,
\qquad
N_{\rm bin}^{\max}
=
\left\lfloor
\log_2\!\left(2^Q\prod_{j=1}^R L_j\right)
\right\rfloor .
\]
For the architecture used in this work, $Q=1$, $R=2$, and $L_1=L_2=8$, giving $N_{\rm bin}^{\max}=7$. 
For fixed cutoff $L$, the representable problem size therefore grows linearly with the number of qumodes, $N_{\rm bin}^{\max}=Q+R\log_2 L$ when $L$ is a power of two. 
For example, a ten-mode memory with one transmon and $L=8$ would in principle encode up to $31$ binary variables, while $L=16$ would encode up to $41$ binary variables. 
If the total hybrid dimension is not an exact power of two, the unused basis states can be assigned large penalty energies or excluded from the decoding map.

The cost of this compression is shifted from physical qubit count to multimode control and photon-number readout. 
For the natural $R$-mode extension of the ECD--rotation ansatz, a depth-$N_d$ circuit contains approximately $RN_d$ ECD--rotation blocks and $4RN_d$ real variational parameters, so the control cost grows linearly with the number of active qumodes at fixed depth and cutoff. 
For independent photon loss, the leading loss contribution is also additive across occupied modes; thus, for comparable mode lifetimes and photon occupations, resonator-loss errors are expected to grow approximately linearly with the number of active modes. \cite{Huang2025fast}
This estimate does not include coherent multimode crosstalk, inherited nonlinearities, calibration errors, or transmon-mediated errors, which remain hardware-dependent.

This scaling picture is consistent with recent multimode cQED demonstrations: Huang \textit{et al.} demonstrated control of a ten-mode bosonic memory with storage-mode lifetimes of the same order, $T_1\simeq0.6$--$1.3~{\rm ms}$, and introduced multimode Fock-state encoding protocols whose pulse count scales with the total number of photons addressed across the modes. \cite{Huang2025fast}
These results suggest that the one-qubit/two-qumode examples studied here can be extended to larger multimode devices, while motivating the photon-loss and coherence estimates discussed below.

Our simulations primarily considered an idealized setting to isolate the expressive advantages of ECD-VQE, though Appendix~\ref{app: noise} examined the impact of photon loss in the qumodes. 
For ECD circuits, the relevant hardware parameter is not the photon-loss rate $\kappa$ alone, but the dimensionless loss accumulated during the circuit, approximately $\kappa T_{\rm circ}$, or equivalently the per-block quantity $\kappa\tau$ when a noise channel is applied after each ECD--rotation block. 
Here $\kappa=1/T_{1,\mathrm{cav}}$ is the photon-loss rate of the resonator and $T_{\rm circ}$ is the total circuit execution time. 
This distinction is important because experimentally demonstrated ECD operations are fast compared with present microwave-cavity lifetimes; for example, Eickbusch \textit{et al.} reported a cavity lifetime $T_{1,\mathrm{cav}}=436\mu{\rm s}$ and demonstrated ECD-based state-preparation sequences with microsecond-scale durations. \cite{Eickbusch2022}

Using these experimental timescales, the circuit depths studied here are realistic with respect to resonator photon loss. 
For the two-qumode ansatz used in this work, each variational layer contains two ECD--rotation blocks, so $N_d=5$ and $N_d=10$ correspond to approximately 10 and 20 ECD blocks, respectively. 
A conservative estimate based on the experimentally demonstrated binomial-code sequences of Eickbusch \textit{et al.} gives an effective timescale of about $0.65\mu{\rm s}$ per ECD block, leading to circuit durations of order $6.5\mu{\rm s}$ for $N_d=5$ and $13\mu{\rm s}$ for $N_d=10$. 
For $T_{1,\mathrm{cav}}=436\mu{\rm s}$, these durations correspond to cumulative photon-loss parameters $\kappa T_{\rm circ}\sim1.5\times10^{-2}$ and $\sim3.0\times10^{-2}$, respectively. 
These values are below the regime where our photon-loss simulations begin to show substantial degradation, while more optimistic pulse-limited estimates give even smaller values. 
The detailed conversion between ECD gate times, circuit depth, and $\kappa T_{\rm circ}$ is given in Appendix~\ref{app: noise}. 
Thus, with respect to resonator photon loss alone, both $N_d=5$ and $N_d=10$ lie within realistic present-day ECD timing budgets.

Decoherence originating from the transmon is largely mitigated through weak dispersive coupling and fast control pulses, \cite{Li2025cascaded,Huang2025fast} as previously demonstrated for ECD-based ansatzes. \cite{You2024Crosstalk}
Quantum error correction applied to higher Fock states can, in principle, suppress photon loss errors, \cite{Noh2020encoding,Wu2023optimalencodingof,dutta2025noise} though achieving this in near-term hardware remains challenging. 
Nonetheless, recent progress in error-corrected logical qubits using bosonic resonators beyond break-even thresholds, \cite{Sivak2023real} and extensions to encoded qudit spaces, \cite{brock2025quantum} highlights a promising pathway for robust hybrid qubit--qumode computation.

In the near term, quantum error mitigation (QEM) provides a practical alternative to full QEC. 
Several QEM protocols have been developed specifically for photon-loss processes in qumodes, \cite{Su2021errormitigationnear,Taylor2024quantum,Mills2024mitigating,teo2025linear}
and integrating these techniques with ECD-VQE is a natural and valuable direction for future work. 
More broadly, the ability to encode large-scale QUBO problems, including those arising in quantum chemistry such as active-space selection, into compact hybrid circuits highlights the potential of qubit-qumode platforms for constrained optimization in early fault-tolerant quantum computing.


\section*{Acknowledgements}

The authors acknowledge partial support from the National Science Foundation Engines Development Award: Advancing Quantum Technologies (CT) under Award Number 2302908. 
VSB acknowledges partial support from the NSF Center for Quantum Dynamics on Modular Quantum Devices (CQD-MQD) under grant number 2124511. 
CW acknowledges partial support from the Department of Energy (Grant No. DE-SC0025521).


\section*{Code and Data Availability}

The Python code and data for the optimizations can be found at\\ \href{https://github.com/CQDMQD/codes_qumode_qubo}{https://github.com/CQDMQD/codes\_qumode\_qubo}.

\appendix

\section*{Appendix A: Effects of qumode noise} \label{app: noise}


\begin{figure}[b!]

\includegraphics[width=0.9\columnwidth]{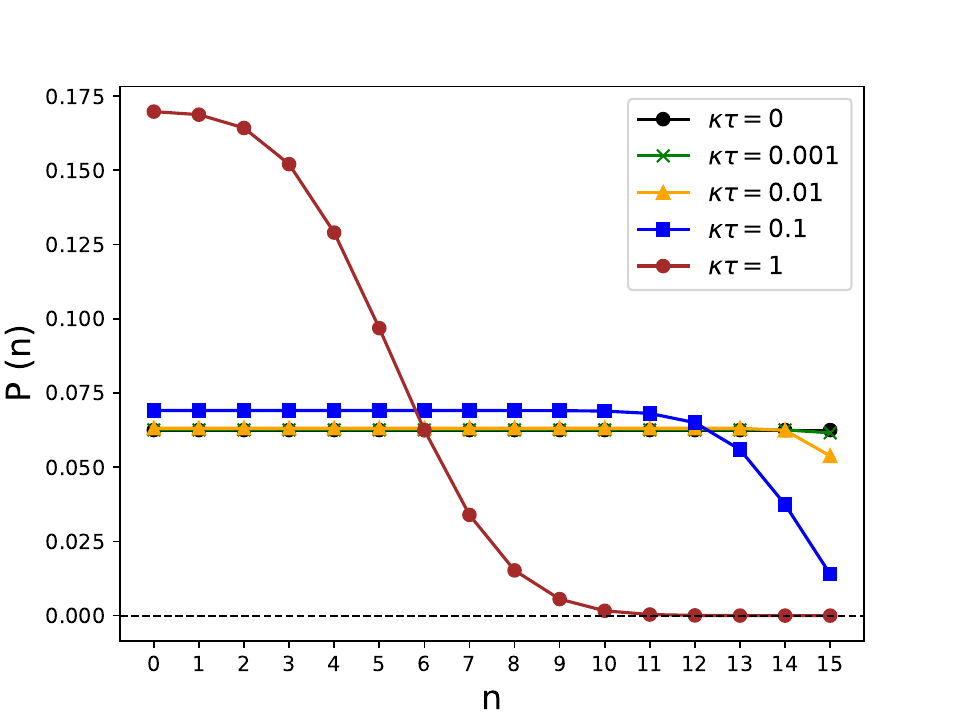}

\caption{
    Effect of amplitude damping as defined in \Eq{\ref{eq: photon_loss_kraus}} on the probabilities $P (n)$ of finding the qumode state $\ket{n}$.  
    The initial state $\rho$ is in an equal superposition state with a Fock cutoff of $L = 16$, which then gets modified due to \Eq{\ref{eq: kraus_amp_damp}} before simulating the photon number measurements.
}
\label{fig: equal_state_qumode_L16_ad}
\end{figure}


\begin{figure*}[t!]

\includegraphics[width=0.9\textwidth]{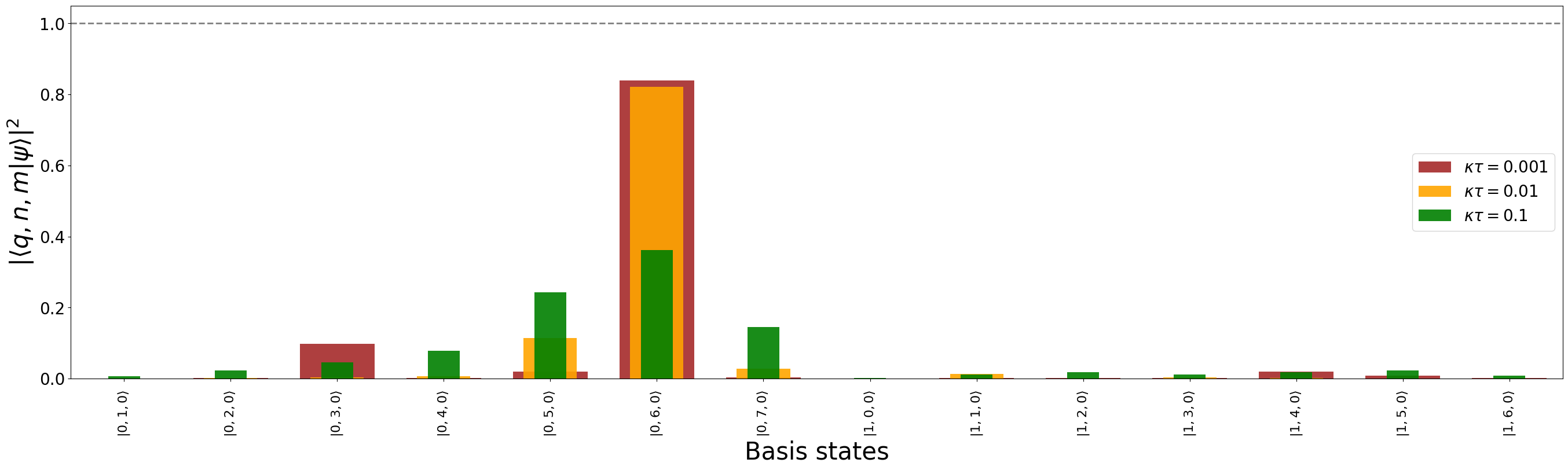}

\caption{
    Effect of qumode amplitude damping as defined in \Eq{\ref{eq: kraus_amp_damp_multimode}} on the photon number probabilities during the ECD-VQE optimization for the BKP problem defined in \Eq{\ref{eq: bkp_q7_example_ham}}.
    The results are for circuit depth $N_d = 5$ with number of iterations = 80. 
    The histograms are split into two parts for better readability of the basis states.  
    The histograms for different noise parameters $\kappa \tau$ are plotted with different widths for better distinguishability. 
    The corresponding noiseless results are shown in \Fig{\ref{fig: bkp_probs_nd5_niters}}.
}
\label{fig: bkp_probs_nd5_niters80_ad}
\end{figure*}


The implementations of the qumode circuits, followed by photon number measurements discussed in this work, are affected by noise in a realistic hardware setup. 
The dominant noise source for microwave resonators in cQED is amplitude damping via photon loss, represented by the following rate equation \cite{liu2026hybrid}:
\begin{equation}
\frac{d }{ d \tau } \: \braket{\BN{}}
= - \kappa \braket{\BN{}},
\end{equation}
where $\tau$ is time and $\kappa$ is the photon loss rate.
The transformation of a qumode density matrix $\rho$ due to the amplitude damping quantum channel can be represented by the Kraus operator formalism, as follows: \cite{NielsenChuang}
\begin{equation} \label{eq: kraus_amp_damp}
\tilde{\rho}
= \sum_{j = 0}^{L - 1} \: K_j \: \rho \: K_j^\dagger,
\end{equation}
where $L$ is the Fock cutoff for the qumode and the Kraus operators are defined, as follows: \cite{Michael2016new}
\begin{equation} \label{eq: photon_loss_kraus}
K_j 
= \sqrt{ \frac{ ( 1 - e^{- \kappa \tau } )^j }{j!} } \: 
e^{- \frac{ \kappa \tau }{2} \BN{} } \: \BA{}^j.
\end{equation}
Due to the truncated expression of \Eq{\ref{eq: kraus_amp_damp}}, the $K_0$ operator must also be modified, as follows: 
\begin{equation}
\tilde{K}_0 
= \Big( \EYE 
- \sum_{j = 1}^{L - 1} \: K_j^\dagger  K_j \Big)^{1/2},
\end{equation}
so that the transformation remains trace-preserving. \cite{liu2026hybrid}
The modified photon number probabilities for a qumode can now be written as,
$ P (n) = \text{Tr} ( \KetBra{n}{n} \tilde{ \rho } ) $.
As an example, we show how amplitude damping, defined according to \Eq{\ref{eq: kraus_amp_damp}}, affects photon number measurements for an initial qumode state where each Fock basis state has equal amplitudes in \Fig{\ref{fig: equal_state_qumode_L16_ad}}.

We discuss how photon loss in the qumodes will affect the ECD-VQE optimizations by applying them to the BKP example, discussed in \Sec{\ref{sec: bkp_example}}.
The density matrix $\rho$ transformation from the noise channel can be represented as 
\begin{equation} \label{eq: kraus_amp_damp_multimode}
\tilde{\rho}
= \sum_{j = 0}^{L_1 - 1} \sum_{k = 0}^{L_2 - 1} \: 
\big( \EYE \otimes K_j \otimes K_k \big)  
\: \rho \: 
\big( \EYE \otimes K_j^\dagger \otimes K_k^\dagger \big),
\end{equation}
where we have assumed the photon loss rate $\kappa$ is the same for each of the qumodes and ignored the noise on the qubit.
We assume that the noise channels affect the quantum state after each block of ECD with qubit rotation block $U_{ER}$ defined in \Eq{\ref{eq: single_ecd_rot}} is applied, i.e., the time parameter $\tau$ in \Eq{\ref{eq: photon_loss_kraus}} will be the circuit execution time for each $U_{ER}$ block. 
After each unitary block is applied, we can represent the updated density matrix as, 
$ \rho \leftarrow U_{ER} \: \rho \: U_{ER}^\dagger $, which then undergoes the noise channel transformation as defined in \Eq{\ref{eq: kraus_amp_damp_multimode}}.
We show the effects of photon loss on the ECD-VQE optimization for the BKP problem in \Eq{\ref{eq: bkp_q7_example_ham}} in \Fig{\ref{fig: bkp_probs_nd5_niters80_ad}}, where we plotted the photon number probabilities after 80 iterations. 
For the noiseless case, 80 iterations are enough to get a resolved peak, as shown in \Fig{\ref{fig: bkp_probs_nd5_niters}}. 
This is also the case up to $\kappa \tau = 0.01$, where the optimization can also resolve the correct solution $\ket{0, 6, 0}$ with certainty. 
However, the optimization performance starts to deteriorate around $\kappa \tau = 0.1$. 
Thus, \Fig{\ref{fig: bkp_probs_nd5_niters80_ad}} gives us an estimate of how the photon loss rate must relate to the implementation time for qubit-qumode on a real quantum device, which is represented by the $\kappa \tau < 0.1$ regime. 

We can connect this dimensionless noise parameter to experimentally demonstrated ECD timescales. \cite{Eickbusch2022}
For a qumode with photon lifetime $T_{1,\mathrm{cav}}$, the photon-loss rate is
\begin{equation}
    \kappa = \frac{1}{T_{1,\mathrm{cav}}}, 
    \qquad
    \kappa \tau = \frac{\tau}{T_{1,\mathrm{cav}}}.
\end{equation}
In the ECD experiment of Eickbusch \textit{et al.}, the cavity lifetime was $T_{1,\mathrm{cav}}=436~\mu{\rm s}$, giving
\begin{equation}
    \kappa \simeq 2.29\times10^3~{\rm s}^{-1},
    \qquad
    \frac{\kappa}{2\pi}\simeq 365~{\rm Hz}.
\end{equation}
In the pulse-constrained regime, the ECD sequence duration has the lower bound
\begin{equation}
    T_{\rm ECD}^{\rm min}
    =
    2t_q + 4t_D,
\end{equation}
where $t_q=24~{\rm ns}$ is the qubit-pulse duration and $t_D=44~{\rm ns}$ is the oscillator-displacement-pulse duration. 
This gives
\begin{equation}
    T_{\rm ECD}^{\rm min}
    =
    2(24~{\rm ns}) + 4(44~{\rm ns})
    =
    224~{\rm ns}
\end{equation}
per ECD block in the most optimistic pulse-limited estimate.

For the two-qumode ansatz considered here, each variational layer contains two ECD--rotation blocks, so that
\begin{equation}
    N_{\rm ECD}\simeq 2N_d .
\end{equation}
Thus, $N_d=5$ and $N_d=10$ correspond roughly to $N_{\rm ECD}=10$ and $N_{\rm ECD}=20$ ECD blocks, respectively. 
Using the pulse-limited timing estimate gives
\begin{equation}
    T_{\rm circ}^{\rm min}(N_d=5)
    \simeq
    10 \times 224~{\rm ns}
    =
    2.24~\mu{\rm s},
\end{equation}
and
\begin{equation}
    T_{\rm circ}^{\rm min}(N_d=10)
    \simeq
    20 \times 224~{\rm ns}
    =
    4.48~\mu{\rm s}.
\end{equation}
For $T_{1,\mathrm{cav}}=436~\mu{\rm s}$, these circuit durations correspond to cumulative photon-loss parameters
\begin{equation}
    \kappa T_{\rm circ}^{\rm min}(N_d=5)
    \simeq
    5.1\times10^{-3},
\end{equation}
and
\begin{equation}
    \kappa T_{\rm circ}^{\rm min}(N_d=10)
    \simeq
    1.0\times10^{-2}.
\end{equation}

A more conservative estimate can be obtained from the experimentally demonstrated binomial-code state preparations by Eickbusch \textit{et al.}, which used ECD depths up to $N_d=5$ with an average pulse duration of $3.27~\mu{\rm s}$. 
Interpreting this as an approximate experimental timescale of $3.27/5\simeq0.65~\mu{\rm s}$ per ECD block gives
\begin{equation}
    T_{\rm circ}(N_d=5)
    \sim
    10 \times 0.65~\mu{\rm s}
    =
    6.5~\mu{\rm s},
\end{equation}
and
\begin{equation}
    T_{\rm circ}(N_d=10)
    \sim
    20 \times 0.65~\mu{\rm s}
    =
    13~\mu{\rm s}.
\end{equation}
The corresponding cumulative photon-loss parameters are
\begin{equation}
    \kappa T_{\rm circ}(N_d=5)
    \sim
    \frac{6.5~\mu{\rm s}}{436~\mu{\rm s}}
    \simeq
    1.5\times10^{-2},
\end{equation}
and
\begin{equation}
    \kappa T_{\rm circ}(N_d=10)
    \sim
    \frac{13~\mu{\rm s}}{436~\mu{\rm s}}
    \simeq
    3.0\times10^{-2}.
\end{equation}
Equivalently, the conservative per-block value is $\kappa\tau\sim0.65/436\simeq1.5\times10^{-3}$, which is below the $\kappa\tau=0.01$ value for which the correct solution remains resolved in \Fig{\ref{fig: bkp_probs_nd5_niters80_ad}}. 
The cumulative photon-loss estimates for both $N_d=5$ and $N_d=10$ are also below the $\kappa\tau\sim0.1$ regime where the photon-loss simulations begin to show substantial degradation. 
This comparison suggests that the circuit depths studied in this work are realistic with respect to resonator photon loss on current ECD hardware.

\bibliography{QuOpt}


\end{document}